\RequirePackage[mathlines]{lineno} 
\documentclass[a4paper,11pt]{article}

\usepackage{jheppub} 
\usepackage{overpic}
\usepackage{amssymb}
\usepackage{amsmath}
\usepackage[T1]{fontenc} 
\usepackage{amsmath}
\usepackage{graphicx}
\usepackage{subfigure}
\usepackage{epstopdf}
\usepackage{rotating}

\usepackage{threeparttable}
\usepackage{multirow}
\usepackage{subfigure}
\usepackage{float}
\usepackage{verbatim}

\let\oldequation\equation
\let\oldendequation\endequation
\renewenvironment{equation}
  {\linenomathNonumbers\oldequation}
  {\oldendequation\endlinenomath}

\begin{document}


\title{\bf \boldmath \texorpdfstring{
Measurement of the branching fraction of $D_s^+\to\tau^+\nu_\tau$ via $\tau^+\to\mu^+\nu_\mu\bar \nu_\tau$}{}
}

\collaborationImg{\includegraphics[height=4cm,angle=0 ]{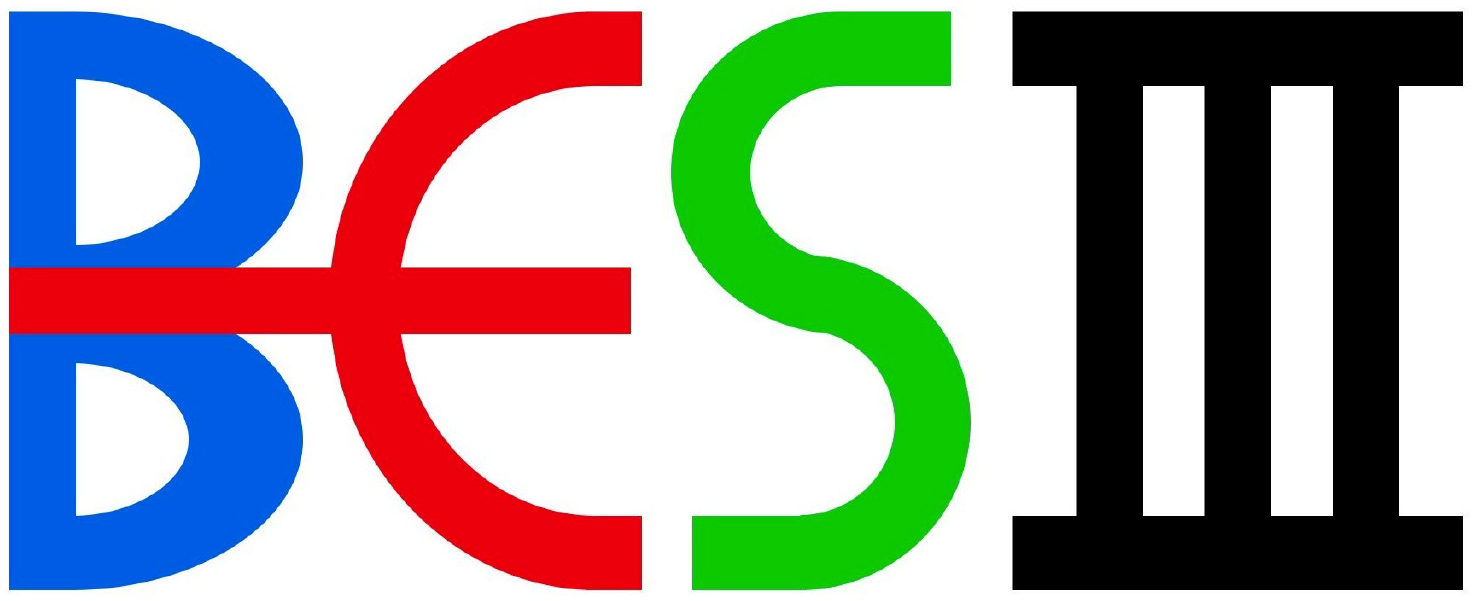}}
\collaboration{BESIII Collaboration}


\author{
M.~Ablikim$^{1}$, M.~N.~Achasov$^{13,b}$, P.~Adlarson$^{73}$, R.~Aliberti$^{34}$, A.~Amoroso$^{72A,72C}$, M.~R.~An$^{38}$, Q.~An$^{69,56}$, Y.~Bai$^{55}$, O.~Bakina$^{35}$, I.~Balossino$^{29A}$
% ...
\\
\vspace{0.2cm}
(BESIII Collaboration)\\
\vspace{0.2cm} {\it
$^{1}$ Institute of High Energy Physics, Beijing 100049, People's Republic of China\\
% ...
$^{13}$ G.I. Budker Institute of Nuclear Physics SB RAS (BINP), Novosibirsk 630090, Russia\\
% ...
$^{29}$ INFN Sezione di Ferrara, (A)INFN Sezione di Ferrara, I-44122, Ferrara, Italy; (B)University of Ferrara,  I-44122, Ferrara, Italy\\
% ...
$^{34}$ Johannes Gutenberg University of Mainz, Johann-Joachim-Becher-Weg 45, D-55099 Mainz, Germany\\
$^{35}$ Joint Institute for Nuclear Research, 141980 Dubna, Moscow region, Russia\\
% ...
$^{38}$ Liaoning Normal University, Dalian 116029, People's Republic of China\\
% ...
$^{55}$ Southeast University, Nanjing 211100, People's Republic of China\\
$^{56}$ State Key Laboratory of Particle Detection and Electronics, Beijing 100049, Hefei 230026, People's Republic of China\\
% ...
$^{69}$ University of Science and Technology of China, Hefei 230026, People's Republic of China\\
% ...
$^{72}$ University of Turin and INFN, (A)University of Turin, I-10125, Turin, Italy; (B)University of Eastern Piedmont, I-15121, Alessandria, Italy; (C)INFN, I-10125, Turin, Italy\\
$^{73}$ Uppsala University, Box 516, SE-75120 Uppsala, Sweden\\
% ...
$^{b}$ Also at the Novosibirsk State University, Novosibirsk, 630090, Russia\\
% ...
}
}

\abstract{
Utilizing $7.33~\mathrm{fb}^{-1}$ of $e^+e^-$ collision data taken at the center-of-mass energies of 4.128, 4.157, 4.178, 4.189, 4.199, 4.209, 4.219, and 4.226\,GeV with the BESIII detector,
the branching fraction of the leptonic decay $D_s^+\to\tau^+\nu_\tau$ via $\tau^+\to\mu^+\nu_\mu\bar \nu_\tau$
is measured to be $\mathcal{B}_{D_s^+\to\tau^+\nu_\tau}=(5.37\pm0.17_{\rm stat}\pm0.15_{\rm syst})\%$.
Combining this branching fraction with the world averages of the measurements of the masses of $\tau^+$ and $D_s^+$ as well as the lifetime of $D_s^+$,
we extract the product of the decay constant of $D_s^+$ and the $c\to s$ Cabibbo-Kobayashi-Maskawa matrix element to be $f_{D_s^+}|V_{cs}|=(246.7\pm3.9_{\rm stat}\pm3.6_{\rm syst})~\mathrm{MeV}$.
Taking $|V_{cs}|$ from a global fit in the standard model we obtain $f_{D_s^+}=(253.4\pm4.0_{\rm stat}\pm3.7_{\rm syst})$\,MeV.  Conversely, taking $f_{D_s^+}$ from lattice quantum chromodynamics calculations, we obtain $|V_{cs}| = 0.987\pm0.016_{\rm stat}\pm0.014_{\rm syst}$.
}

\maketitle
\flushbottom

\section{INTRODUCTION}

Leptonic decays offer an ideal laboratory for studying strong and weak interaction effects in the charmed meson system.
In the standard model~(SM) of particle physics, the $D^+_s$ meson decays into $\ell^+\nu_\ell$~($\ell=e$, $\mu$ or $\tau$) via annihilation mediated by a virtual $W^+$ boson. Throughout this paper, the inclusion of charge conjugate channels is always implied.
The partial width of $D^+_s\to \ell^+\nu_\ell$ at  lowest order
can be related to the $D^+_{s}$ decay constant $f_{D^+_{s}}$ via~\cite{decayrate}
\begin{equation}
\Gamma_{D^+_{s}\to\ell^+\nu_\ell}=\frac{G_F^2}{8\pi}|V_{cs}|^2
f^2_{D^+_{s}}
m_\ell^2 m_{D^+_{s}} \left (1-\frac{m_\ell^2}{m_{D^+_{s}}^2} \right )^2,
\label{decayratedstolnu_draft}
\end{equation}
where $G_F$ is the Fermi coupling constant,
$|V_{cs}|$ is the $c\to s$ Cabibbo-Kobayashi-Maskawa (CKM) matrix element,
$m_\ell$ is the mass of the lepton, and $m_{D^+_{s}}$ is the mass of  the $D^+_{s}$ meson.
Extraction of $f_{D^+_{s}}$ in experiments is important for testing various theoretical calculations based on different approaches~\cite{FLab2018,LQCD,etm2015,ukqcd2017,ukqcd2015,FLAG2021,chen2014,becirevic2013,wang2015}.
In recent years, the precision of calculations of $f_{D^+_s}$ based on Lattice Quantum Chromodynamics~(LQCD) has reached a level of 0.2\%~\cite{FLAG2021}, and much progress has been achieved in the experimental studies of  $D^+_s\to \ell^+\nu_\ell$ decays
by the CLEO~\cite{cleo2009,cleo2009a,cleo2009b}, BaBar~\cite{babar2010}, Belle~\cite{belle2013}, and
BESIII~\cite{bes2016,bes2019,hajime2021,tauvyue,huijingevv,xiechen} collaborations.
Based on the average of the branching fractions (BFs) reported by these experiments, one can derive $f_{D^+_s}$ with a precision of 1.0\%.
Precise and intensive estimations of $f_{D^+_s}$ are still desirable to test theoretical calculations with higher precision. 
Improved measurements of $f_{D_s} \times\left|V_{\rm cs}\right|$ are therefore important for testing the unitarity of the CKM matrix~\cite{haibo2021} with higher sensitivity.

In the SM, the ratio of the BFs of $D^+_s\to \tau^+\nu_\tau$
and $D^+_s\to \mu^+\nu_\mu$ can be written as
\begin{equation}
\mathcal R_{\tau/\mu}=\frac{\mathcal B_{D^+_s\to \tau^+\nu_\tau}}{\mathcal B_{D^+_s\to \mu^+\nu_\mu}}=\frac{m_{\tau^+}^2(1-\frac{m_{\tau^+}^2}{m_{D^+_s}^2})^2}{m_{\mu^+}^2(1-\frac{m_{\mu^+}^2}{m_{D^+_s}^2})^2},
\end{equation}
which only depends on the charged lepton and $D_s^{+}$ meson masses.
Inserting the world averages of $m_\tau$, $m_\mu$, and $m_{D_s}$~\cite{PDG2022} in the above equation gives $\mathcal R_{\tau/\mu}=9.75\pm0.01$.
Measurements of the BFs of $D^+_s\to \ell^+\nu_\ell$ allow this ratio to be determined experimentally and provide an important test of $\tau-\mu$ lepton flavor universality.

In this paper, we present a measurement of the BF of $D_s^+\to\tau^+\nu_\tau$
via the decay of $\tau^+\to\mu^+\nu_\mu\bar \nu_\tau$, by analyzing $7.33~\mathrm{fb}^{-1}$ of $e^+e^-$ collision data
taken at the center-of-mass energies $\sqrt s=$ 4.128~GeV, 4.157~GeV, 4.178~GeV, 4.189~GeV, 4.199~GeV, 4.209~GeV, 4.219~GeV, and 4.226~GeV~\cite{BESIII:2015zbz,ref_emc_energy,ref_emc_energy2} with the BESIII detector~\cite{Ablikim:2009aa}.
Following previous measurements, we have not corrected the BF of $D^+_s\to \tau^+\nu_\tau$ by the effect of radiative photons since their uncertainties can be considered individually later,
details of which are reviewed in ``{\it Leptonic Decays of Charged Pseudoscalar Mesons}'' by the Particle Data Group (PDG)~\cite{PDG2022}.
Based on this measurement, we determine $f_{D_s^{+}} \times |V_{\rm cs}|$ with an improved accuracy, and test $\tau-\mu$ lepton flavor universality with $D_s^{+} \rightarrow \ell^{+} \nu_{\ell}$ decays.

\section{BESIII DETECTOR AND MONTE CARLO SIMULATION}

The BESIII detector~\cite{Ablikim:2009aa} records symmetric $e^+e^-$ collisions
provided by the BEPCII storage ring~\cite{Yu:IPAC2016-TUYA01} in the center-of-mass energy range from 2.00 to 4.95 GeV, with a peak luminosity of $1\times10^{33}$~cm$^{-2}$s$^{-1}$ achieved at $\sqrt{s} = 3.77$~GeV.
BESIII has collected large data samples in this energy region~\cite{white_paper}. The cylindrical core of the BESIII detector covers 93\% of the full solid angle and consists of a helium-based
 multilayer drift chamber~(MDC), a plastic scintillator time-of-flight
system~(TOF), and a CsI(Tl) electromagnetic calorimeter~(EMC),
which are all enclosed in a superconducting solenoidal magnet
providing a 1.0~T magnetic field~\cite{youbes}. The solenoid is supported by an
octagonal flux-return yoke with modules of resistive plate muon counters
(MUC) interleaved with steel.
The charged-particle momentum resolution at $1~{\rm GeV}/c$ is
$0.5\%$, and specific ionization energy loss
d$E$/d$x$ resolution is $6\%$ for electrons
from Bhabha scattering. The EMC measures photon energies with a
resolution of $2.5\%$ ($5\%$) at $1$~GeV in the barrel (end-cap)
region. The time resolution in the TOF barrel region is 68~ps. The end-cap TOF
system was upgraded in 2015 using multi-gap resistive plate chamber
technology, providing a time resolution of
60~ps~\cite{etofa,etofb,etofc}.
Approximately 83\% of the data used here was collected
after this upgrade.

Simulated data samples, namely inclusive MC samples, produced with a {\sc
geant4}-based~\cite{geant4} Monte Carlo (MC) package, which
includes the geometric description of the BESIII detector and the
detector response, are used to determine detection efficiencies
and to estimate backgrounds. The simulation models the beam-energy spread and
initial-state radiation (ISR) in the $e^+e^-$
annihilations with the generator {\sc
kkmc}~\cite{ref:kkmc1,ref:kkmc2}.
In the simulation, the production of open-charm
processes directly produced via $e^+e^-$ annihilations are modeled with the generator {\sc conexc}~\cite{ref:conexc},
and their subsequent decays are modeled by {\sc evtgen}~\cite{ref:evtgen1,ref:evtgen2} with
known BFs from the Particle Data Group~\cite{PDG2022}.
The ISR production of vector charmonium (-like) states
and the continuum processes are incorporated in {\sc kkmc}~\cite{ref:kkmc1,ref:kkmc2}.
The remaining unknown charmonium decays
are modeled with {\sc lundcharm}~\cite{ref:lundcharm1,ref:lundcharm2}. Final-state radiation
from charged final-state particles is incorporated using the {\sc
photos} package~\cite{photos}.
The input cross section line shape of $e^+e^-\to D_s^{*\pm}D_s^\mp$ is based on the cross section measurement in the energy range from threshold to 4.7~GeV.

\section{ANALYSIS METHOD}

In $e^+e^-$ collisions with data taken at the center-of-mass energies between 4.128 and 4.226~GeV,
the $D_s^\pm$ mesons are produced mainly via the $e^+e^-\to D_s^{*\pm}D_s^\mp\to \gamma(\pi^0)D_s^+ D_s^-$ process.
For our analysis we adopt the double-tag~(DT) method pioneered by the MARK III Collaboration~\cite{DTmethod1}.
The $D_s^-$ meson, when fully reconstructed via any hadronic decay mode, is referred to as the single-tag (ST) $D_s^-$ meson.
Events in which the transition $\gamma(\pi^0)$ from the $D_s^{*+}$ meson and the leptonic decay of $D_s^+\to \tau^+\nu_\tau$ are reconstructed, in addition to the ST $D^-_s$ meson, are denoted as DT events.
The BF of $D^+_s\to \tau^+\nu_\tau$ is determined by
\begin{equation}
\mathcal B^j_{D_s^+\to\tau^+\nu_\tau}=\frac{N_{\rm DT}^j/  \epsilon^j_{\rm DT} }{\mathcal B_{\tau^+\to\mu^+\nu_\mu\bar \nu_\tau} \cdot  N_{\rm ST}^{j} / \epsilon^j_{\rm ST} },
\label{eq1}
\end{equation}
where $N_{\rm DT}^j$ and $N_{\rm ST}^j$ are the yields of the DT events and ST $D^-_s$ mesons in data, respectively;
and $\epsilon_{\rm DT}^j$ and $\epsilon_{\rm ST}^j$ are the efficiencies of the DT events and ST $D^-_s$ mesons estimated with MC simulation, respectively.
Here, $\epsilon_{\rm DT}^j$, which includes the efficiency of simultaneously finding the tag side, the transition $\gamma(\pi^0)$ and $D_s^+\to\tau^+\nu_\tau$ as well as the BF of $D^{*+}_s\to \gamma(\pi^0)D^+_s$,
$\mathcal B_{\tau^+\to\mu^+\nu_\mu\bar \nu_\tau}$ is the BF of $\tau^+\to\mu^+\nu_\mu\bar \nu_\tau$ and $j$ denotes the ST mode. The weighted mean method~\cite{Schmelling:1994pz} is utilized to calculate the final BF, taking into account the statistical and tag mode dependent uncertainty as discussed later.

\section{\texorpdfstring{SINGLE-TAG CANDIDATES}{}}

To reconstruct  ST $D^-_s$ candidates, we use the fourteen hadronic decay modes
$D^-_s\to K^+K^-\pi^-$, $K^+K^-\pi^-\pi^0$, $K^0_SK^-$,
$K^0_SK^-\pi^0$,
$K^0_SK^0_S\pi^-$,
$K^0_SK^+\pi^-\pi^-$,
$K^0_SK^-\pi^+\pi^-$,
$\pi^+\pi^-\pi^-$,
$\eta_{\gamma\gamma}\pi^-$,
$\eta_{\pi^0\pi^+\pi^-}\pi^-$,
$\eta^\prime_{\eta_{\gamma\gamma}\pi^+\pi^-}\pi^-$,
$\eta^\prime_{\gamma\rho^0}\pi^-$,
$\eta_{\gamma\gamma}\rho^-$, and
$\eta_{\pi^+\pi^-\pi^0}\rho^-$.
Throughout this paper, $\rho$ denotes $\rho(770)$ and the subscripts of $\eta^{(\prime)}$ denote individual decay modes adopted for the $\eta^{(\prime)}$ reconstruction.

In selecting $K^\pm$, $\pi^\pm$, $K^0_S$, $\gamma$, $\pi^0$, and $\eta$ candidates, we use the same selection criteria as those adopted in our previous studies~\cite{bes2019,bes3_etaev,bes3_gev}.
For each good charged track, the polar angle ($\theta$) with respect to the beam direction is required to be
within the MDC acceptance $|\!\cos\theta|<0.93$, where $\theta$ is defined with respect to the $z$ axis, which is the symmetry axis of the MDC.
The distance of its closest approach relative to the interaction point is required to be within 10.0~cm along the beam direction~($|V_{z}|$)
and within 1.0~cm in the plane transverse to the beam direction~($|V_{xy}|$).
Particle identification (PID) for good charged tracks combines the measurements of
the d$E$/d$x$ in the MDC and the flight time in the TOF to form probabilities $\mathcal{L}(h) (h = K, \pi)$
for each hadron $(h)$ hypothesis. The charged tracks are assigned as kaons or pions if their probabilities satisfy $\mathcal{L}(K) > \mathcal{L}(\pi)$ and $\mathcal{L}(\pi) > L(K)$, respectively.

$K_S^0$ candidates are reconstructed via $K^0_S\to \pi^+\pi^-$ decays.
The two charged pions are required to satisfy $|V_{z}|<20$ cm and $|\!\cos\theta|<0.93$.
They are assumed to be $\pi^+\pi^-$ without particle identification (PID) requirements and their invariant mass is required to be within $(0.486, 0.510)$~GeV$/c^2$.
The distance from the $K_S^0$ decay vertex to the interaction point is required to be greater than twice the vertex resolution.

Photon candidates are selected by using the information measured by the EMC and are required to satisfy the following criteria.
The energy of each shower in the barrel (end-cap) region of the EMC~\cite{Ablikim:2009aa} is required to be greater than 25 (50) MeV.
To suppress backgrounds associated with charged tracks,
the angle between the shower position and the closest intersection point of any charged track with the EMC inner surface, projected from the interaction point, must be greater than 10 degrees.
To suppress electronic noise and energy deposits unrelated
to the event of interest,
any candidate shower is required to start within $[0, 700]$ ns from the event start time.

$\pi^0$ and $\eta_{\gamma\gamma}$ candidates are formed from $\gamma\gamma$ pairs with invariant masses lying in the mass intervals $(0.115,\,0.150)$ and $(0.50,\,0.57)$\,GeV$/c^{2}$, respectively.
To improve momentum resolution, each selected $\gamma\gamma$ pair is subjected to a  kinematic fit that constrains their invariant mass to the known $\pi^{0}$ or $\eta$ mass~\cite{PDG2022}.
In order to form $\rho^{+(0)}$, $\eta_{\pi^0\pi^+\pi^-}$, $\eta^\prime_{\eta\pi^+\pi^-}$, and $\eta^\prime_{\gamma\rho^0}$ candidates,
the invariant masses of the $\pi^+\pi^{0(-)}$, $\pi^0\pi^+\pi^-$, $\eta\pi^+\pi^-$, and  $\gamma\rho^0$ combinations are required to lie
within the mass intervals of $(0.57,\,0.97)~\mathrm{GeV}/c^2$, $(0.53,\,0.57)~\mathrm{GeV}/c^2$,  $(0.946,\,0.970)~\mathrm{GeV}/c^2$ and $(0.940,\,0.976)~\mathrm{GeV}/c^2$, respectively.
In addition, the energy of the photon from the $\eta^\prime_{\gamma\rho^0}$ decay is required to be greater than 0.1~GeV.

Soft pions from $D^{*+}$ decays are suppressed by requiring the momentum of any pion which is not from $K_S^0$, $\eta$, or $\eta^\prime$ to be greater than 0.1~GeV/$c$. In order to reject the peaking background from $D^-_s\to K^0_S\pi^-$ decays in the selection of $D^-_s\to \pi^+\pi^-\pi^-$ STs, the invariant mass of any $\pi^+\pi^-$ combination is required to lie outside the mass window of $(0.468, 0.528)$~GeV/$c^2$.

\begin{figure}[htbp]
  \centering
  \includegraphics[width=0.8\textwidth] {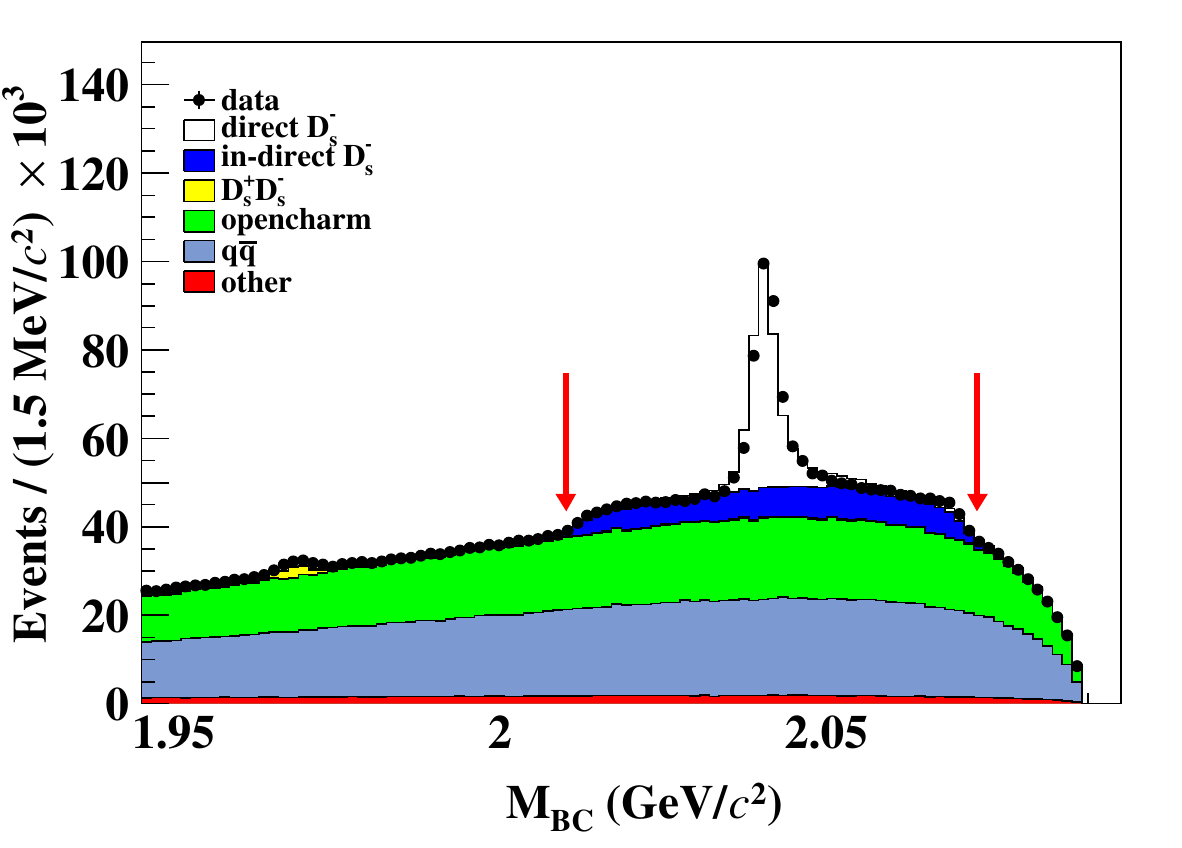}
  \caption{
  The $M_{\rm BC}$ distributions of the ST $D^-_s$ candidates in data and inclusive MC samples at 4.178 GeV. The candidates between the two red arrows are retained for further analysis. 
  }
  \label{fig:diffmbc_ds}
\end{figure}

The backgrounds from non-$D_s^{\pm}D^{*\mp}_s$ processes are suppressed by using the beam-constrained mass of the ST $D_s^-$ candidate
defined as
\begin{equation}
M_{\rm BC}\equiv  \sqrt{E_{\rm beam}^2-|\vec{p}_{\rm ST}|^2},
\end{equation}
where $E_{\rm beam}$ is the beam energy~($\sqrt{s}/2$) and
$\vec{p}_{\rm ST}$ is the momentum of the ST $D_s^-$ candidate in the $e^+e^-$ rest frame.
Figure~\ref{fig:diffmbc_ds} shows the $M_{\rm BC}$ distribution of the ST candidates at 4.178~GeV.
The $M_{\rm BC}$ value is required to be within $(2.010,\,2.061+i\times0.003)~\mathrm{GeV}/c^2$,
where $i$ takes the value 0, 3, 4, 5, 6, 7, 8, 9 for the energy points 4.128, 4.157, 4.178, 4.189, 4.199, 4.209, 4.219, 4.226, respectively.
This requirement retains most of the $D_s^-$ and $D_s^+$ mesons from $e^+ e^- \to D_s^{*\mp}D_s^{\pm}$ production.

If there are multiple candidates present  per tag mode per charge,
only the one with the $D_s^-$ recoil mass
\begin{equation}
M_{\rm rec} \equiv \sqrt{ \left (\sqrt s - \sqrt{|\vec p_{\rm ST}|^2+m^2_{D^-_s}} \right )^2
-|\vec p_{\rm ST}|^2}
\end{equation}
closest to the $D_s^{*+}$ nominal mass~\cite{PDG2022} is kept for further analysis.

\begin{figure}[htbp]
\centering
\includegraphics[width=0.99\textwidth]{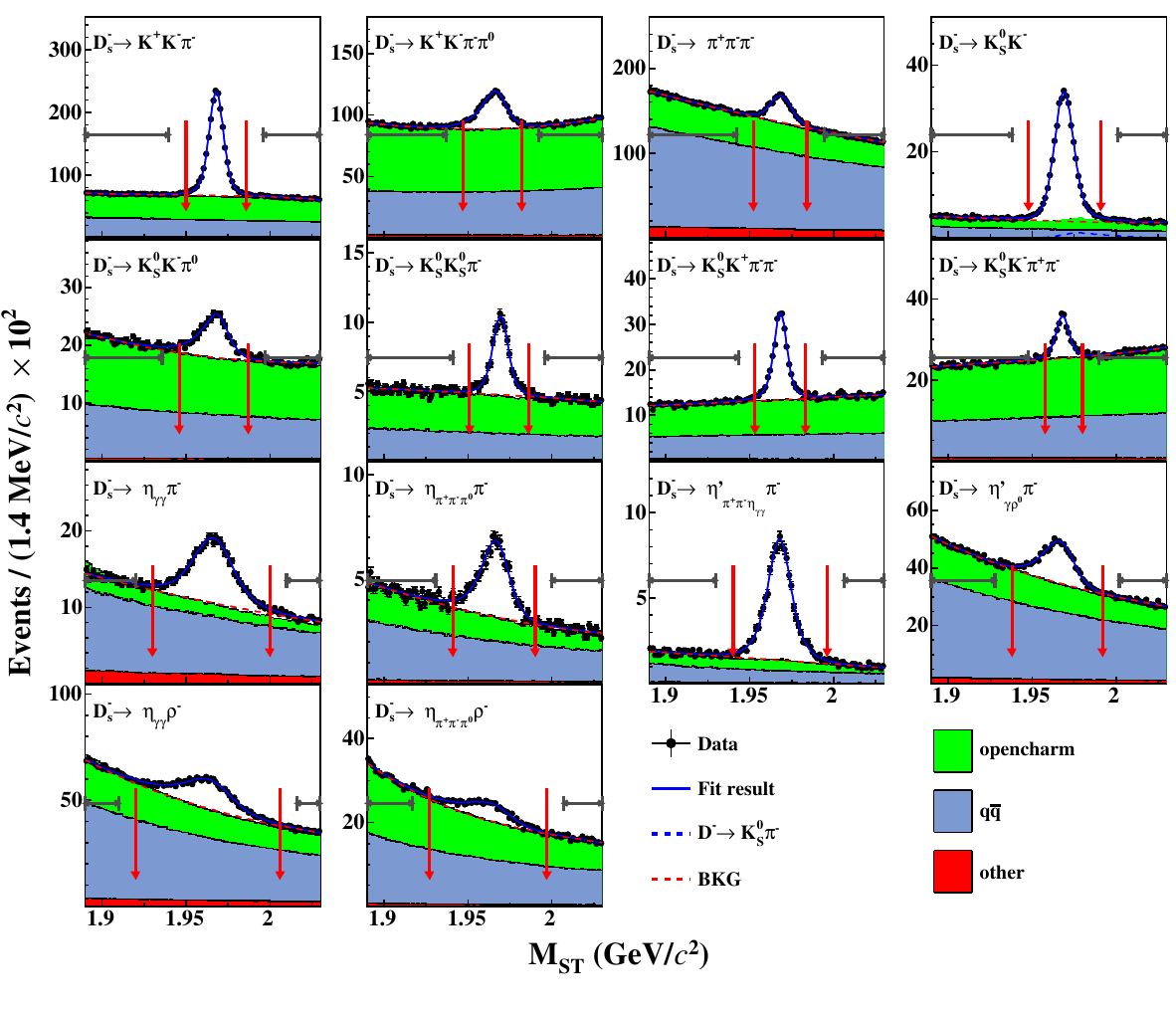}
\caption{
The fits to the $M_{\rm ST}$ distributions of the surviving ST $D^-_s$ candidates for each tag mode.
The points with error bars denote the data sample at $\sqrt s=$ 4.178~GeV.
The blue solid curves represent the best fit results.
The red dashed curves represent the fitted backgrounds.
For the $D^-_s\to K_S^0K^-$ tag mode, the blue dotted curve is the peaking background from $D^-\to K_S^0\pi^-$.
In each figure, the range within the two arrows indicate the chosen $M_{\rm ST}$ signal regions and the brown line segments indicate the sideband regions.
}
\label{fig:stfit}
\end{figure}

The distributions of the invariant masses ($M_{\rm ST}$) of the accepted ST candidates from data for each tag mode are shown in Fig.~\ref{fig:stfit}.
The yields of ST $D^-_s$ mesons reconstructed in each tag mode are determined from fits to their individual $M_{\rm ST}$ distributions.
In the fits, the signal is described by the simulated shape convolved with a Gaussian function that represents the resolution difference between data and simulation.
In the fit to the $D_s^-\to K_S^0K^-$ tag mode,
the shape of the peaking background  $D^-\to K^0_S\pi^-$ is modeled by the simulated shape convolved with the same Gaussian resolution function as used for the signal shape and its size is left free. The fraction of the $D^-\to K^0_S\pi^-$ over $D_s^-\to K_S^0 K^-$ yields is about 2.0\%.
The combinatorial background is described by a first to third order Chebychev function, which is validated by analyzing the inclusive MC sample.
Figure~\ref{fig:stfit} shows the fit results for the data sample at $\sqrt s=$ 4.178~GeV.
In each sub-figure, the red arrows show the chosen $M_{\rm ST}$ signal regions.
The candidates located in these signal regions are retained for further analysis.
Based on simulation, the $e^+e^-\to(\gamma_{\rm ISR})D_s^+D_s^-$ process is found to contribute about (0.7-1.1)\% in the fitted number of ST $D^-_s$ mesons for each tag mode.  The reported yields have this contribution subtracted.
The efficiencies of reconstructing ST $D^-_s$ mesons~($N_{\rm ST}$) are estimated by analyzing the inclusive MC sample in the same way as real data.

The second and third columns of Table~\ref{DT_sum} summarize the yields of ST $D^-_s$ mesons ($N_{\rm ST}$) for each tag mode obtained in data and the corresponding detection efficiencies ($\epsilon_{\rm ST}$), respectively.
In this table, the $N_{\rm ST}$ quantities are obtained by summing over all energy points,
and the $\epsilon_{\rm ST}$ quantities are obtained by weighting the
corresponding yields of ST $D^-_s$ mesons in data at each energy points.

\section{\texorpdfstring{DOUBLE-TAG CANDIDATES}{}}

The $D_s^+\to\tau^+\nu_\tau$ candidates are selected in the system recoiling against the ST $D_s^-$ mesons
via the decay of $\tau^+\to \mu^+\nu_\mu\bar \nu_\tau$
by using the residual neutral showers and charged tracks which have not been used in the ST selection.
As the detection efficiencies and background levels do not vary greatly with $\sqrt{s}$, the analysis combines the samples over all the energy points.

\begin{figure}[htbp]
  \centering
  \includegraphics[width=0.9\textwidth]{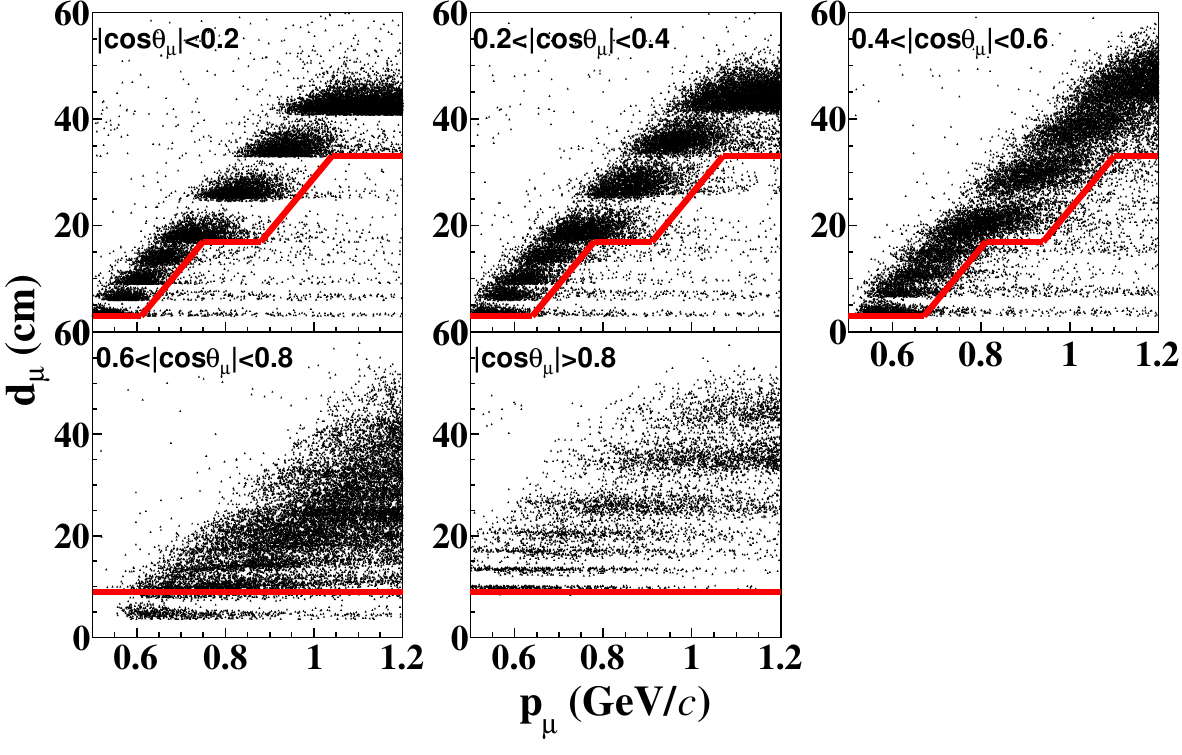}
  \caption{
    The distributions of
  $d_{\mu}$ vs. $p_{\mu}$ in different $|\!\cos\theta_{\mu}|$ regions of $e^+e^-\to \gamma \mu^+\mu^-$ candidates in data.
  The regions above the red line are retained for further analysis.}
  \label{fig:muid_cos}
  \end{figure}

Excluding the daughter particles originating from the tag side, only one good charged track is allowed in each DT candidate and
its charge must be opposite to that of the tag-side decay.
The deposited energy of muon candidates in the EMC is required to be within $(0.0,~0.3)$~GeV.
To separate muons from hadrons, the muon candidates must  have momenta greater than $0.5$~GeV/$c$, and fulfill requirements on the muon travelling length in the MUC ($d_\mu$) with dependence of momentum ($p_\mu$) and flight direction ($\cos\theta_{\mu}$) in the MUC~\cite{bes2019} as shown in Table~\ref{tab:muonid} and Fig.~\ref{fig:muid_cos} based on the control sample of $e^+e^-\to \gamma\mu^+\mu^-$.

\begin{table}[hbtp]\centering
\caption{Identification criteria for muon candidates.}
\scalebox{0.9}{
\begin{tabular}{lcl} \hline\hline
\multicolumn{1}{c}{$|\!\cos\theta_{\mu}|$ }& $p_{\mu}$ (GeV/$c$) & \multicolumn{1}{c}{$d_{\mu}$ (cm)} \\ \hline
          &    $(0.50,~0.61)$   &$>3.0$        \\
          &    $(0.61,~0.75)$   &$>100.0\times p_{\mu}-58.0$ \\
(0.0,~0.2) &   $(0.75,~0.88)$   &$>17.0$ \\
          &    $(0.88,~1.04)$   &$>100.0\times p_{\mu}-71.0$ \\
          &    $(1.04,~1.20)$   &$>33.0$ \\ \hline
          &    $(0.50,~0.64)$   &$>3.0$        \\
          &    $(0.64,~0.78)$   &$>100.0\times p_{\mu}-61.0$ \\
(0.2,~0.4) &   $(0.78,~0.91)$   &$>17.0$ \\
          &    $(0.91,~1.07)$   &$>100.0\times p_{\mu}-74.0$ \\
          &    $(1.07,~1.20)$   &$>33.0$ \\ \hline
          &    $(0.50,~0.67)$   &$>3.0$        \\
          &    $(0.67,~0.81)$   &$>100.0\times p_{\mu}-64.0$ \\
(0.4,~0.6) &   $(0.81,~0.94)$   &$>17.0$ \\
          &    $(0.94,~1.10)$   &$>100.0\times p_{\mu}-77.0$ \\
          &    $(1.10,~1.20)$   &$>33.0$ \\ \hline
(0.6,~0.8) &                    &$> 9.0$ \\ \hline
(0.8,~0.93)&                    &$> 9.0$ \\ \hline\hline
\end{tabular}
}
\label{tab:muonid}
\end{table}

To select the $D_s^+\to\tau^+\nu_\tau$ signals and the transition $\gamma(\pi^0)$ from $D_s^{*+}$, we define two kinematic variables: the energy difference
\begin{equation}
\Delta E \equiv \sqrt s-E_{\rm ST}-E_{\rm miss}-E_{\gamma(\pi^0)},
\end{equation}
where $E_{\rm miss}$ is defined as $\sqrt{|\vec{p}_{\rm miss}|^2+m_{D_s^+}^2}$ with
$\vec{p}_{\rm miss} \equiv -\vec{p}_{\rm ST}-\vec{p}_{\gamma(\pi^0)}$,
and the missing mass squared of the neutrinos
\begin{equation}
M_{3\nu}^2 \equiv\left (\sqrt s-\Sigma_k E_k\right )^2 -|\Sigma_k \vec{p}_k|^2,
\end{equation}
in which $E_k$ and $\vec p_k$ are the energy and momentum of ST $D^-_s$, transition $\gamma(\pi^0)$, or $\mu^+$, respectively.
All $\gamma$ and $\pi^0$ candidates that have not been used in tag selection are looped over.
If there are multiple $\gamma$ or $\pi^0$ combinations satisfying the selection criteria, we choose the one leading to the minimum $|\Delta E|$.

\begin{figure}[htbp]\centering
  \includegraphics[width=0.9\textwidth]{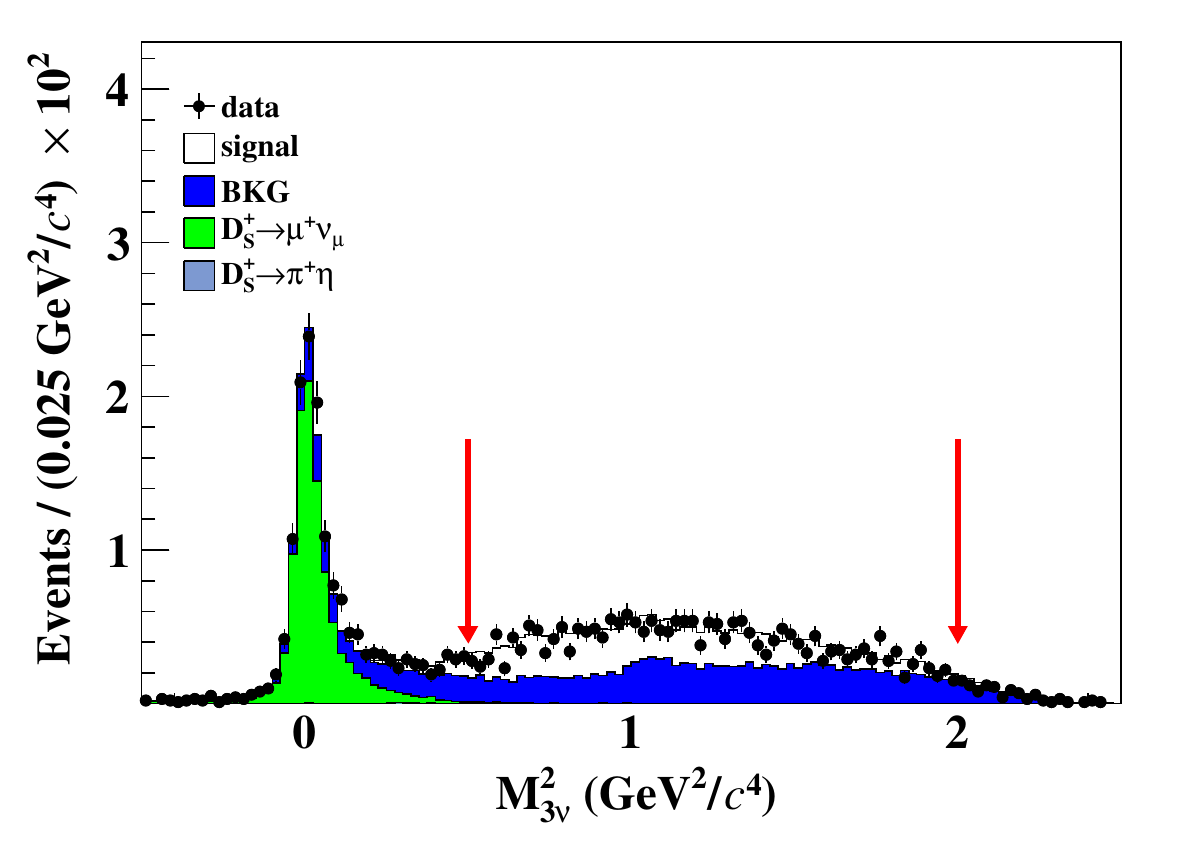}
  \caption{
    The $M_{3\nu}^2$ distributions of accepted candidates in data and the inclusive MC samples with the $E_{\rm extra~\gamma}^{\rm tot}<$0.4 GeV requirement. Candidates with $M_{3\nu}^2$ within the two red arrows are retained for further analysis.
  }
  \label{fig:m21}
\end{figure}

To suppress the backgrounds from  $D_s^+ \rightarrow \mu^+ \nu_{\mu}$ and $D_s^+ \rightarrow \eta \pi^+$ decays, which peak in the $M_{3\nu}^2$ distribution around 0 and 0.3~GeV$^2$/$c^4$, respectively, the value of $M_{3\nu}^2$ is required to be within (0.5,~2.0)~GeV$^2$/$c^4$  as shown in Fig.~\ref{fig:m21}.

\section{BRANCHING FRACTION DETERMINATION}

\begin{figure*}[htpb]\centering
\includegraphics[width=1.0\textwidth]{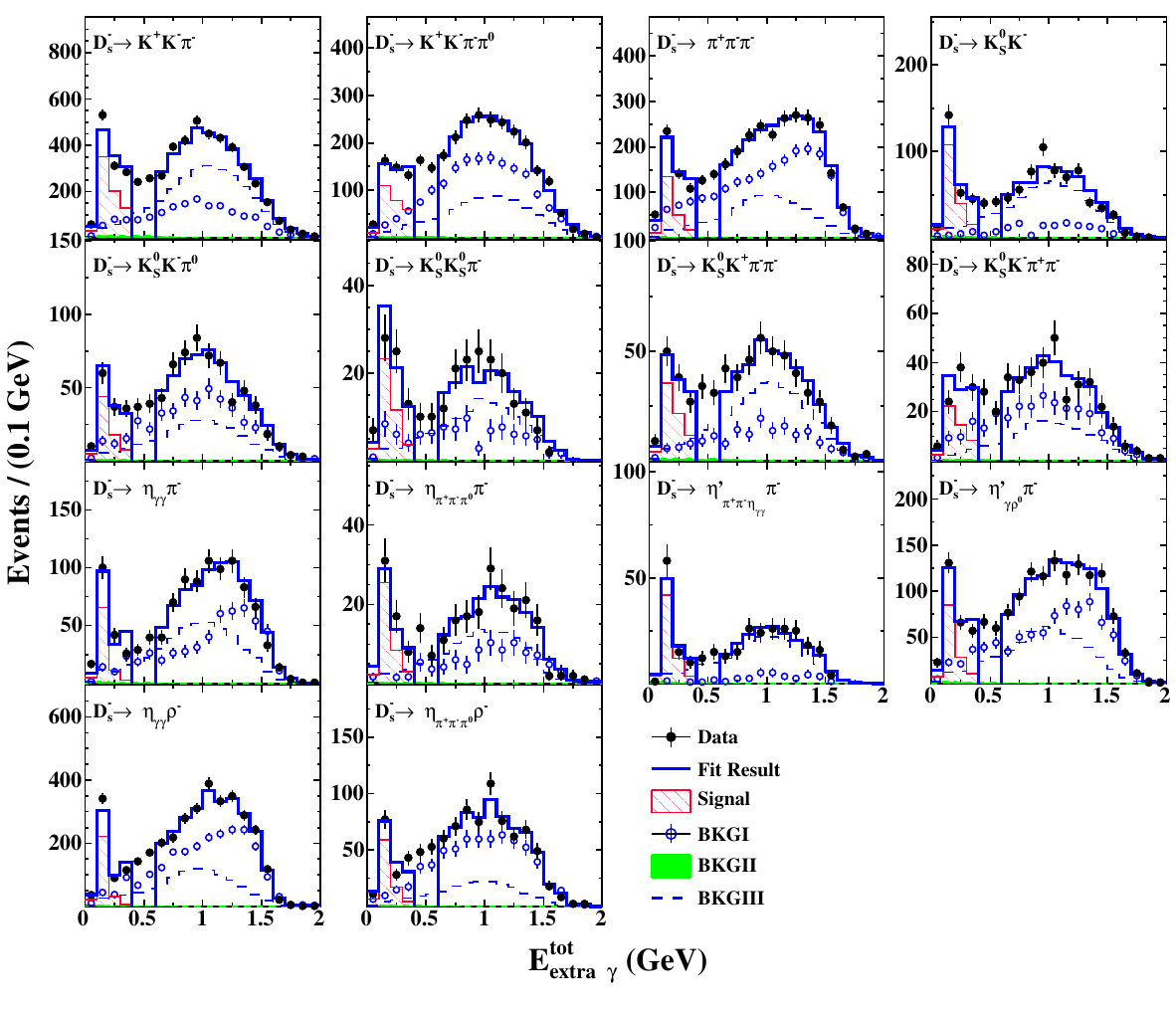}
\caption{
The distributions of $E^{\rm tot}_{\rm extra\,\gamma}$ of the DT candidates for $D^+_s\to \tau^+\nu_\tau$ with $\tau^+\to \mu^+\nu_\mu \bar \nu_\tau$.
Black points with error bars are the combined data sample.
Solid blue histograms denote the resutlts. Filled pink shadows, open circles with error bars, filled green histograms, and dashed blue histograms
are Signal, BKGI, BKGII, and BKGIII, respectively.
The area to the left of the red arrow denotes the signal region.
}
\label{fit_etot_data_4178}
\end{figure*}

Following Refs.~\cite{huijingevv,cleo_dstotaunu_enunu1,cleo_dstotaunu_enunu2}, we discriminate signal from background by using the variable $E^{\rm tot}_{\rm extra\,\gamma}$.
It is defined as the total energy of the good isolated EMC showers which have not been used in tag selection.
The  distributions of $E^{\rm tot}_{\rm extra\,\gamma}$ of the accepted DT candidates in data are shown in Fig.~\ref{fit_etot_data_4178}.

Study of the inclusive MC sample shows that the background events can be divided into three categories: BKGI, BKGII, and BKGIII. The BKGI component corresponds to events with an incorrectly reconstructed ST $D_s^{-}$. The BKGII component corresponds to events with a correctly reconstructed ST $D_s^-$ and $D_s^+ \to K_L^0 \mu^+\nu_\mu$, in which the $K_L^0$ meson passes through the detector without undergoing decay or significant interaction.
The BKGIII component consists of events with a correctly reconstructed ST $D_s^{-}$ and a $D_s^{+}$ decaying to any other background final state apart from $K_L^0 \mu^{+} \nu_\mu$,

The DT signal yield is extracted by analyzing the $E^{\rm tot}_{\rm extra\,\gamma}$ distribution as shown in Fig.~\ref{fit_etot_data_4178}.
To minimize the effect of the imperfect signal shape, we adopt an extrapolation technique following Refs.~\cite{huijingevv,cleo_dstotaunu_enunu1,cleo_dstotaunu_enunu2}. 
A bin maximum likelihood fit is performed on the events with $E^{\rm tot}_{\rm extra\,\gamma}>0.6$~GeV, where the signal is negligible, and 
the sizes and shapes of BKGI and BKGII are fixed.
The signal DT yield is obtained by subtracting the yields of BKGI, BKGII, and BKGIII from the yield of all events~($N^j_{\rm tot}$) in the $E_{\rm extra \gamma}^{\rm tot}$ signal region. 
In the $D_s^*$ rest frame, the transition photon has a monochromatic energy of $139$~MeV. When evaluated in the laboratory rest frame, the $D_s^*$ momentum causes a smearing of $\pm 15$~MeV on the photon energy. After further considering the resolution effect, we define  $E_{\rm extra \gamma}^{\rm tot}<0.4$~GeV as the signal region. Details of BKGI, BKGII, and BKGIII are given below.
 
The shape of the BKGI component is derived using the data DT events situated in the corresponding $M_{\rm ST}$ sideband regions.
The $M_{\rm ST}$ sideband regions are indicated inside the brown line segments in Fig.~\ref{fig:stfit}.
For tag modes with neutrals, the remaining contamination from signal in sideband regions is subtracted.
The size of this component is fixed at
$f^j_1 \cdot N_{\rm Class}^{{\rm I}~j}$,
where $f^j_1$ is the sideband scale factor, 
defined as the ratio of the numbers of background events in the $M_{\rm ST}$ sideband and signal ranges.
The $f^j_1$ value is obtained by fitting the $M_{\rm ST}$ distribution from the inclusive MC sample after imposing the DT requirements.
$N_{\rm Class}^{{\rm I}~j}$ is obtained by counting events in the $E^{\rm tot}_{\rm extra\,\gamma}$ signal region in data.

The shape of the BKGII component is modeled by the simulated events corrected by a 2D data-MC difference for the $K_L^0$ detector response. The correction factors are obtained by using a control sample of $D^0 \to K_L^0 \pi^+ \pi^-$ decays from 2.93~fb$^{-1}$ of $e^+e^-$ collision data collected at $\sqrt s=3.773$~GeV~\cite{luminosity_3773,luminosity_3773a}. The yield of this component is fixed at $N_{\rm Class}^{{\rm II}~j}$, which is calculated by taking the probability not to reconstruct the $K^0_L$ meson from MC simulation and assuming the BF of $D_s^+ \to K^0 \mu^+\nu_\mu$ decays to be the same as the corresponding decay mode involving electrons~\cite{PDG2022}.

The shape of the BKGIII component is estimated from the inclusive MC sample.
The MC simulation shows that the leading six $D^+_s$ non-peaking background components are
$D_s^+ \to \eta \mu^+ \nu_\mu$ (36.0\%),
$D_s^+ \to \eta \pi^+ \pi^0$ (11.4\%),
$D_s^+ \to \pi^+ \pi^0 \nu_\tau \bar{\nu_\tau}$ (2.5\%),
$D_s^+ \to \phi \pi^+$ (2.5\%),
$D_s^+ \to \eta^\prime \pi^+$ (2.5\%),
and
$D_s^+ \to \phi \mu^+ \nu_\mu$ (2.0\%), where the numbers shown in parentheses are their proportional contribution to the total BKGIII in the full $E^{\rm tot}_{\rm extra\,\gamma}$ region.
The yield of this component is represented by $f^j_2 \cdot N_{\rm Class}^{{\rm III}~j}$, where $f^j_2$ 
 is the extrapolation factor, defined as the ratio of the numbers of BKGIII events between $E^{\rm tot}_{\rm extra\,\gamma}<0.4$ GeV and $E^{\rm tot}_{\rm extra\,\gamma}>0.6$ GeV derived from the inclusive MC sample.
The $N_{\rm Class}^{{\rm III}~j}$ is obtained from the fit with $E^{\rm tot}_{\rm extra\,\gamma}>0.6$ GeV.

Finally, the signal DT yield in data is obtained by
\begin{equation}
  \label{cal_event}
  N^j_{\rm DT}=N^j_{\rm tot}-f^j_1 \cdot N_{\rm Class}^{{\rm I}~j}-N_{\rm Class}^{{\rm II}~j}-f^j_2 \cdot N_{\rm Class}^{{\rm III}~j}.
\end{equation}

The efficiencies of detecting DT events ($\epsilon^j_{\rm DT}$) are estimated by using the signal MC samples of $e^+e^-\to D_s^\mp D^{*\pm}_s$ with the $D_s^-$ meson decaying to the tag mode and $D_s^+\to \tau^+\nu_\tau$ with $\tau^+\to\mu^+\nu_\mu\bar \nu_\tau$.
All numbers discussed above are summarized in Table~\ref{DT_sum}. For each tag mode,
inserting the individual values of $N^j_{\rm ST}$, $\epsilon^j_{\rm ST}$, $N^j_{\rm DT}$, and $\epsilon^j_{\rm DT}$ in Eq.~\ref{eq1} gives the corresponding BF.  The systematic uncertainties in the BF measurement are estimated in the next section. The obtained BFs are summarized in the last column of Table~\ref{DT_sum}.

\begin{sidewaystable}[htbp]
  \caption{The fitted yields of ST $D^-_s$ mesons in data ($N^j_{\rm ST}$);
  the efficiencies of detecting ST $D^-_s$ mesons ($\epsilon^j_{\rm ST}$) and DT events ($\epsilon^j_{\rm DT}$) for each tag mode;
  the number of total DT events ($N^j_{\rm tot}$);
  the sideband scale factor of BKGI ($f^j_1$);
  the extrapolation factor of BKGIII ($f^j_2$);
  the BKGI yield within $E^{\rm tot}_{\rm extra\,\gamma}<0.4$ GeV ($N^{{\rm I}~j}_{\rm Class}$);
  the BKGII yield within $E^{\rm tot}_{\rm extra\,\gamma}<0.4$ GeV ($N^{{\rm II}~j}_{\rm Class}$);
  the BKGIII yield within $E^{\rm tot}_{\rm extra\,\gamma}>0.4$ GeV ($N^{{\rm III}~j}_{\rm Class}$);
  and the net numbers of DT events ($N^j_{\rm DT}$). 
  For the obtained $\mathcal{B}^j_{D_s^+ \to \tau^+ \nu_{\tau}}$, the first, second, and third uncertainties are the statistical, tag-mode dependent systematic and tag-mode independent systematic, respectively.
  The listed efficiencies do not include the BFs of the sub decays.
  The index $j$ from 1 to 14 represents the tag modes
  $D_{s}^{-}\to K^{+} K^{-}\pi^{-}$,
  $D_{s}^{-}\to K^{+} K^{-}\pi^{-}\pi^{0}$,
  $D_{s}^{-}\to \pi^{+}\pi^{-}\pi^{-}$,
  $D_{s}^{-}\to K_S^{0} K^{-}$,
  $D_{s}^{-}\to K_S^{0} K^{-}\pi^{0}$,
  $D_{s}^{-}\to K_S^{0} K_S^{0}\pi^{-}$,
  $D_{s}^{-}\to K_S^{0} K^{+}\pi^{-}\pi^{-}$,
  $D_{s}^{-}\to K_S^{0} K^{-}\pi^{+}\pi^{-}$,
  $D_{s}^{-}\to \eta_{\gamma\gamma}\pi^{-}$,
  $D_{s}^{-}\to \eta_{\pi^{+}\pi^{-}\pi^{0}}\pi^{-}$,
  $D_{s}^{-}\to \eta\prime_{\pi^{+}\pi^{-}\eta} \pi^{-}$,
  $D_{s}^{-}\to \eta\prime_{\gamma\rho^{0}} \pi^{-}$,
  $D_{s}^{-}\to \eta_{\gamma\gamma}\rho^{-}_{\pi^{-}\pi^{0}}$,
  and $D_{s}^{-}\to \eta_{\pi^{+}\pi^{-}\pi^{0}}\rho^{-}_{\pi^{-}\pi^{0}}$, respectively.
  The $\epsilon^j_{\rm DT}/\epsilon^j_{\rm ST}$ varies within 46\% for different tag modes; this is mainly due to the significantly different signal environments for some tag modes containing low momentum photon and pions in the signal and inclusive MC samples.}
  \label{DT_sum}
    \centering
  \scalebox{0.80}{
  \begin{tabular}{cr@{}lr@{}lr@{}lr@{}lr@{}lr@{}lr@{}lr@{}lr@{}lr@{}lc}
    \hline   \hline
$j$&\multicolumn{2}{c}{$N_{\rm ST}^{j}$ ($\times 10^3$)}&\multicolumn{2}{c}{$\epsilon_{\rm ST}^{j}$ (\%)}&\multicolumn{2}{c}{$\epsilon_{\rm DT}^{j}$ (\%)}&\multicolumn{2}{c}{$N_{\rm tot}^{j}$}&\multicolumn{2}{c}{$f_1^{j}$}&\multicolumn{2}{c}{$N_{\rm Class}^{{\rm I}~j}$}&\multicolumn{2}{c}{$N_{\rm Class}^{{\rm II}~j}$} &\multicolumn{2}{c}{$f_2^j$} & \multicolumn{2}{c}{$N_{\rm Class}^{{\rm III}~j}$} & \multicolumn{2}{c}{$N_{\rm DT}^{j}$} & ${\mathcal B}^{j}_{D^+_s \to \tau^+\nu_\tau}$~(\%)\\ \hline
1&280.7&$\pm$0.9&40.87&$\pm$0.01&12.62&$\pm$0.06&1184.0&$\pm$34.4&0.422&$\pm$0.001&531.0&$\pm$23.0&54.0&$\pm$6.8&0.080&$\pm$0.001&2413.2&$\pm$65.1&713.9&$\pm$36.1&$5.42\pm0.27\pm0.05\pm0.14$\\
2&86.3&$\pm$1.3&11.83&$\pm$0.01&4.61&$\pm$0.04&472.0&$\pm$21.7&0.396&$\pm$0.001&337.7&$\pm$18.4&18.4&$\pm$2.5&0.086&$\pm$0.001&700.7&$\pm$52.1&259.3&$\pm$23.4&$5.08\pm0.46\pm0.13\pm0.13$\\
3&72.7&$\pm$1.4&51.86&$\pm$0.03&16.80&$\pm$0.16&536.0&$\pm$23.2&0.355&$\pm$0.001&671.0&$\pm$25.9&15.8&$\pm$1.9&0.094&$\pm$0.002&706.1&$\pm$52.2&215.6&$\pm$25.4&$6.02\pm0.71\pm0.13\pm0.16$\\
4&62.2&$\pm$0.4&47.37&$\pm$0.03&14.96&$\pm$0.16&251.0&$\pm$15.8&0.672&$\pm$0.009&27.0&$\pm$5.2&13.3&$\pm$1.7&0.093&$\pm$0.002&490.1&$\pm$26.6&173.7&$\pm$16.4&$5.81\pm0.55\pm0.08\pm0.15$\\
5&23.0&$\pm$0.6&17.00&$\pm$0.03&6.66&$\pm$0.11&143.0&$\pm$12.0&0.508&$\pm$0.003&82.5&$\pm$9.1&6.0&$\pm$0.7&0.102&$\pm$0.003&205.0&$\pm$27.5&74.1&$\pm$13.1&$5.42\pm0.96\pm0.18\pm0.14$\\
6&10.4&$\pm$0.2&22.51&$\pm$0.05&7.71&$\pm$0.19&73.0&$\pm$8.5&0.403&$\pm$0.004&48.0&$\pm$6.9&2.3&$\pm$0.3&0.102&$\pm$0.005&97.1&$\pm$13.4&41.4&$\pm$9.1&$7.65\pm1.68\pm0.25\pm0.20$\\
7&29.6&$\pm$0.3&20.98&$\pm$0.03&7.14&$\pm$0.11&124.0&$\pm$11.1&0.336&$\pm$0.002&62.0&$\pm$7.9&6.2&$\pm$0.8&0.089&$\pm$0.003&272.2&$\pm$21.2&72.6&$\pm$11.6&$4.73\pm0.76\pm0.09\pm0.12$\\
8&15.3&$\pm$0.4&18.23&$\pm$0.03&6.26&$\pm$0.14&98.0&$\pm$9.9&0.231&$\pm$0.001&157.0&$\pm$12.5&3.3&$\pm$0.4&0.088&$\pm$0.004&121.9&$\pm$19.4&47.6&$\pm$10.5&$5.96\pm1.31\pm0.21\pm0.16$\\
9&39.6&$\pm$0.8&48.31&$\pm$0.04&16.86&$\pm$0.21&185.0&$\pm$13.6&1.256&$\pm$0.012&40.0&$\pm$6.3&9.8&$\pm$1.1&0.106&$\pm$0.003&376.3&$\pm$34.8&85.2&$\pm$16.2&$4.06\pm0.77\pm0.11\pm0.11$\\
10&11.7&$\pm$0.3&23.31&$\pm$0.05&8.49&$\pm$0.20&56.0&$\pm$7.5&0.604&$\pm$0.009&7.8&$\pm$2.8&2.9&$\pm$0.3&0.094&$\pm$0.004&100.4&$\pm$15.0&39.0&$\pm$7.8&$6.02\pm1.20\pm0.22\pm0.16$\\
11&19.7&$\pm$0.2&25.17&$\pm$0.04&8.82&$\pm$0.16&84.0&$\pm$9.2&0.848&$\pm$0.019&2.0&$\pm$1.4&4.8&$\pm$0.5&0.106&$\pm$0.004&158.3&$\pm$15.0&60.7&$\pm$9.4&$5.78\pm0.89\pm0.15\pm0.15$\\
12&50.1&$\pm$1.0&32.46&$\pm$0.03&11.35&$\pm$0.13&277.0&$\pm$16.6&0.743&$\pm$0.003&115.5&$\pm$10.7&12.1&$\pm$1.5&0.102&$\pm$0.002&455.8&$\pm$39.1&132.4&$\pm$18.9&$4.97\pm0.71\pm0.12\pm0.13$\\
13&80.1&$\pm$2.3&19.92&$\pm$0.01&8.70&$\pm$0.07&581.0&$\pm$24.1&2.315&$\pm$0.012&79.4&$\pm$8.9&26.7&$\pm$3.4&0.112&$\pm$0.002&814.3&$\pm$80.4&279.6&$\pm$33.0&$5.25\pm0.62\pm0.18\pm0.14$\\
14&22.2&$\pm$1.4&9.15&$\pm$0.01&4.11&$\pm$0.06&159.0&$\pm$12.6&1.272&$\pm$0.008&37.7&$\pm$6.1&7.4&$\pm$0.9&0.111&$\pm$0.003&156.9&$\pm$36.4&86.3&$\pm$15.4&$5.70\pm1.01\pm0.39\pm0.15$\\
  \hline \hline
  \end{tabular}
  }
  \end{sidewaystable}

\section{SYSTEMATIC UNCERTAINTIES}
\label{sys}

Sources of the relative systematic uncertainties in the measurement of the BF of $D^+_s\to \tau^+\nu_\tau$ are summarized in Table~\ref{tab:sys_tot} and discussed below.  Note that the DT method means that most uncertainties due to the selection of ST $D^-_s$ candidates cancel.

\subsection{TAG-MODE DEPENDENT SYSTEMATIC UNCERTAINTIES}

Several sources of potential systematic bias are associated with the tag mode, and are hence classified as tag-mode dependent.

The systematic uncertainties on the fitted yields of the ST $D^-_s$ mesons are assessed by
using alternative signal and background shapes.
The alternative signal shapes are obtained by changing the baseline choices derived from inclusive MC sample to those from the signal MC sample.
The alternative background shapes are obtained by varying the order of the nominal Chebychev function by $\pm 1$.
For a given ST mode, the differences in the ratio of the yields of ST $D^-_s$ mesons over the corresponding efficiency for all variations,
and the background fluctuation of the fitted yield of ST $D^-_s$ are re-weighted by the yields of ST $D^-_s$ mesons in various data samples and are added in quadrature.  An additional component to this uncertainty is statistical in nature, and accounts for the contribution of background fluctuations to the fitted yields of ST $D^-_s$ mesons. The effects due to the signal shape, the background shape, and the background fluctuation are 0.08\%, 0.12\%, and 0.46\%, respectively.
The corresponding overall systematic uncertainty from all these sources is assigned to be 0.48\%, which is the quadrature sum of these three terms.

The ST efficiencies obtained from the inclusive MC sample may differ from those estimated with the signal MC events generated with events containing the ST $D_s^-$ and $D_s^+\to\tau^+\nu_\tau$ decays, thereby causing possible tag bias.
The size of this bias is estimated by measuring for each tag  $\varepsilon_{\rm ST}^{D_s^+\to \tau^+\nu_\tau}$, the efficiency in the signal MC sample,  and $\varepsilon_{\rm ST}^{{\rm inclusive}\, D_s^+}$, the efficiency in the inclusive MC sample, and multiplying $(\varepsilon_{\rm ST}^{D_s^+\to \tau^+\nu_\tau}/\varepsilon_{\rm ST}^{{\rm inclusive}\, D_s^+} - 1)$ by the estimated data-MC differences in the tracking and PID efficiencies without any correction, which are 1.0\% for charged pions and kaons, and 2.0\% for  $\pi^0$, $\eta(\gamma\gamma)$ and $K^0_S$ decays.  The resulting numbers are weighted by the ST yields in each tag to yield an overall systematic uncertainty of 0.37\%.

After weighting by the yields of ST $D^-_s$ mesons in each data sample, the uncertainty from the limited MC sample sizes is assigned to be 0.29\%.

\subsection{TAG-MODE INDEPENDENT SYSTEMATIC UNCERTAINTIES}

Systematic uncertainties which do not depend on tag modes are classified as tag-mode independent.

The systematic uncertainties related to the $\mu^+$ tracking and PID efficiencies are investigated by using a control sample of $e^+e^-\to\gamma \mu^+ \mu^-$ decays.
By considering the dependencies of the $\mu^+$ efficiencies on the $\mu$ momentum, polar angle, and different energy points, the difference of $\mu^+$ tracking efficiencies between data and MC simulation is $(-0.32\pm0.18)\%$. After correcting the signal efficiencies to data, the associated systematic uncertainty is assigned to be 0.18\%.
The difference of the $\mu^+$ PID efficiencies between data and MC simulation is found to be $-(11.86\pm0.33)\%$.
A similar large difference in the $\mu^+$ PID efficiency between data and simulation was observed for  $D_s^+\to\mu^+\nu_\mu$ events in previous analyses at BESIII and is understood to arise from imperfections in the simulation of the length of the muon traveling in the MUC~\cite{bes2019}.
After correcting the signal efficiencies to data, the uncertainty 0.33\% is assigned as the corresponding systematic uncertainty.

The efficiency of the $\gamma$ selection is studied by using a control sample of $J/\psi\to\pi^+\pi^-\pi^0$ decays~\cite{geff}, while the $\pi^0$ reconstruction efficiency is studied with a  sample of $e^+e^-\to K^+K^-\pi^+\pi^-\pi^0$ events~\cite{pi0eff}.
The systematic uncertainty of selecting the transition $\gamma$ or $\pi^0$ is estimated to be 1.00\%,
accounting for the relative BFs of $D_s^{*+}\to\gamma D_s^+$ and $D_s^{*+}\to\pi^0D_s^+$~\cite{PDG2022}.

The systematic uncertainty associated with the $M^{2}_{3\nu}$ requirement is assessed by re-performing the measurement with enlarging or shrinking this requirement by $\pm 1$ or $\pm 2$ bin sizes, resulting in 24 variations. Among all variations, the maximum change of BF, $1.75\%$, is taken as the corresponding systematic uncertainty.

The systematic uncertainty associated with the requirement of no extra charged tracks ($N_{\rm extra}^{\rm charge}$) is studied with the DT sample
of $D_s^+ \to \pi^+ \phi(\to K^+K^-)$ and $D_s^+ \to K^+ K^0_S(\to \pi^+\pi^-) $.
The difference of the acceptance efficiencies between data and MC simulation, 0.41\%, is taken as the systematic uncertainty.

The systematic uncertainty in the $E^{\rm tot}_{\rm extra\,\gamma}$ fit has contributions associated with the three classes of background.
The systematic uncertainty arising from the BKGI is estimated by varying the sideband scale factor by $\pm 1\sigma$ and the corresponding change of 0.10\% in the fitted signal yield  is taken as the systematic uncertainty.
The systematic uncertainty arising from the shape of BKGII is assessed by replacing the corrected shape of $E^{\rm tot}_{\rm extra\,\gamma}$ with the uncorrected one and is found to be negligible.
We also change the level of BKGII background by varying  the misidentification rate  by $\pm 1\sigma$ and the BF of $D_{s}^{+}\to K_{L}^0 \mu^+ \nu_{\mu}$ within the measurement uncertainty of  the $D_{s}^{+}\to K_{L}^0 e^+ \nu_{e}$  BF.
The relative difference of the fitted signal yield, 1.39\%, is assigned as the associated systematic uncertainty.
The uncertainty due to the non-peaking shape of BKGIII is estimated by varying the $f_2$ by $\pm 1\sigma$ and the relative components of the leading six background modes~\cite{PDG2022}, and is  assigned to be 0.69\%.
After adding these contributions in quadrature, the uncertainty associated with the $E^{\rm tot}_{\rm extra\,\gamma}$ fit is assigned to be 1.56\%.

The uncertainty on the BF of $\tau^+\to\mu^+\nu_\tau \bar \nu_\tau$ contributes a systematic uncertainty of 0.23\%~\cite{PDG2022}.

\subsection{TOTAL SYSTEMATIC UNCERTAINTIES}

By adding the individual components in quarature, we determine the total tag-mode dependent and independent systematic uncertainties to be 0.67\% and 2.62\%, respectively, and the total relative systematic uncertainty to be 2.70\%.

\begin{table}[htbp]\centering
\caption{ Systematic uncertainties in the BF measurement.}
\scalebox{1.00}{
\begin{tabular}{cc}
\hline
\hline
Source & Uncertainty\,(\%)\\ \hline
ST yield                                 &0.48\\
Tag bias                                         &0.37\\
MC sample size                                    &0.29\\
\hline
   $\mu^+$ tracking                          &0.18         \\
   $\mu^+$ PID                               &0.33         \\
   $\gamma(\pi^0)$ reconstruction                &1.00         \\
   $M_{3\nu}^2$ requirement              &1.75   \\
   $N_{\rm extra}^{\rm charge}$ requirement  &0.41\\
   $E^{\rm tot}_{\rm extra\,\gamma}$ fit      &1.56         \\
   $\mathcal{B}(\tau^+\to \mu^+ \nu_{\mu} \bar \nu_{\tau})$ &0.23\\
\hline
Total                                            &2.70\\
\hline
\hline
\end{tabular}}
\label{tab:sys_tot}
\end{table}

\section{RESULTS}

The measured values  $\mathcal{B}_{D_s^+\to\tau^+ \nu_{\tau}}$ are listed in Table~\ref{DT_sum} for each tag mode.
Weighting each measurement  by the inverse squares of the combined statistical and tag-mode dependent systematic uncertainties yields
\begin{equation}
{\mathcal B}_{D_s^+\to\tau^+\nu_\tau}=(5.37\pm0.17_{\rm stat}\pm0.15_{\rm syst})\%.
\nonumber
\end{equation}
Here, the first uncertainty is statistical, and the second is the quadrature sum of the tag-mode dependent and independent systematic uncertainties.
Using this BF and the world average values of $G_F$, $m_\mu$, $m_{D^+_s}$, and $\tau_{D^+_s}$~\cite{PDG2022} with $\Gamma_{D_s^+\to\tau^+\nu_\tau}=\mathcal B_{D_s^+\to\tau^+\nu_\tau}/\tau_{D_s^+}$, we determine the product of $f_{D_s^+}$ and $|V_{cs}|$ to be
\begin{equation}
f_{D_s^+}|V_{cs}|=(246.7\pm3.9_{\rm stat}\pm3.6_{\rm syst})~{\rm MeV},
\nonumber
\end{equation}
where the  systematic uncertainty is dominated by that of the  measured BF (1.86\%) and the lifetime of $D^+_s$ (0.8\%).
Making use of $|V_{cs}|=0.97349\pm0.00016$ from the global fit in the SM~\cite{PDG2022, ckmfitter}, we obtain
\begin{equation}
f_{D_s^+}=(253.4\pm4.0_{\rm stat}\pm3.7_{\rm syst})~{\rm MeV}.
\nonumber
\end{equation}
Alternatively, utilizing $f_{D_s^+}=(249.9\pm0.5)~\mathrm{MeV}$ from recent LQCD calculations~\cite{FLab2018,LQCD,etm2015,FLAG2021}, we obtain
\begin{equation}
|V_{cs}|=0.987\pm0.016_{\rm stat}\pm0.014_{\rm syst}.
\nonumber
\end{equation}
In the calculation of $|V_{cs}|$, one additional uncertainty (0.2\%) for the input value of $f_{D_s^+}$  is included.
In the determination of $f_{D_s^+}$, however, the uncertainty from the input value $|V_{cs}|$ has negligible effect.
Our value $|V_{cs}|$  agrees with our previous results obtained via
$D\to\bar K\ell^+\nu_\ell$~\cite{bes3_kev,bes3_ksev,bes3_klev,bes3_kmuv},  $D_s^+\to\mu^+\nu_\mu$~\cite{bes2019}, and $D_s^+\to\eta^{(\prime)}e^+\nu_e$ decays~\cite{bes3_etaev}.

\section{SUMMARY}
By analyzing $e^+e^-$ collision data collected 
with a total integrated luminosity of $7.33~\mathrm{fb}^{-1}$
at the center-of-mass energies between 4.128~GeV and 4.226~GeV,
we determine the BF of $D_s^+\to\tau^+\nu_\tau$
via $\tau^+\to\mu^+\nu_\mu\bar \nu_\tau$ to be $(5.37\pm0.17_{\rm stat}\pm0.15_{\rm syst})\%$.
This result is consistent with the previous measurements~\cite{PDG2022}.
Using this BF and the world average values of $G_F$, $m_\mu$, $m_{D^+_s}$, and $\tau_{D^+_s}$~\cite{PDG2022} with $\Gamma_{D_s^+\to\tau^+\nu_\tau}=\mathcal B_{D_s^+\to\tau^+\nu_\tau}/\tau_{D_s^+}$, we determine the product of $f_{D_s^+}$ and $|V_{cs}|$ to be
$f_{D_s^+}|V_{cs}|=(246.7\pm3.9_{\rm stat}\pm3.6_{\rm syst})~{\rm MeV}$.
Combining the BF measured in this work with the $|V_{cs}|$ given by Refs.~\cite{PDG2022, ckmfitter},
we obtain $f_{D^+_s}=(253.4\pm4.0_{\rm stat}\pm3.7_{\rm syst})$~MeV.
Conversely, combining this BF with the $f_{D^+_s}$ calculated by  LQCD~\cite{FLab2018,LQCD,etm2015,FLAG2021},
we determine $|V_{cs}|=0.987\pm0.016_{\rm stat}\pm0.014_{\rm syst}$.
Combining with the BF of $D^+_s \to \mu^+\nu_\mu$~\cite{PDG2022},
we obtain
$\mathcal R_{\tau/\mu}=9.89\pm0.50$, which is consistent with the expectation based on lepton flavor universality. 

\begin{figure}[htbp]\centering
\includegraphics[width=0.8\textwidth]{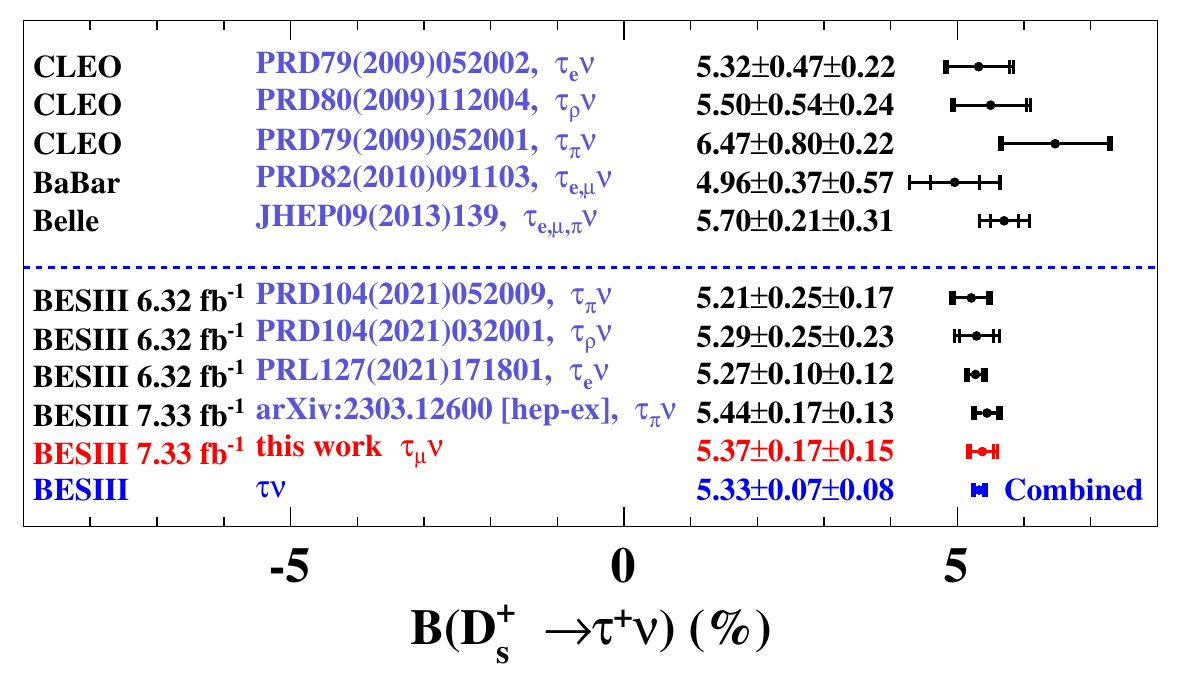}
\caption{Comparison of the BFs measured in this work with previous
measurements, where the inner error bar is the statistical uncertainty and the outer is the combined statistical and systematic uncertainty. The last line is the BESIII combined result which does not include the BESIII result in Ref.~\cite{hajime2021}.}
\label{fig:combf}
\end{figure}

\begin{figure}[htbp]\centering
\includegraphics[width=0.7\textwidth]{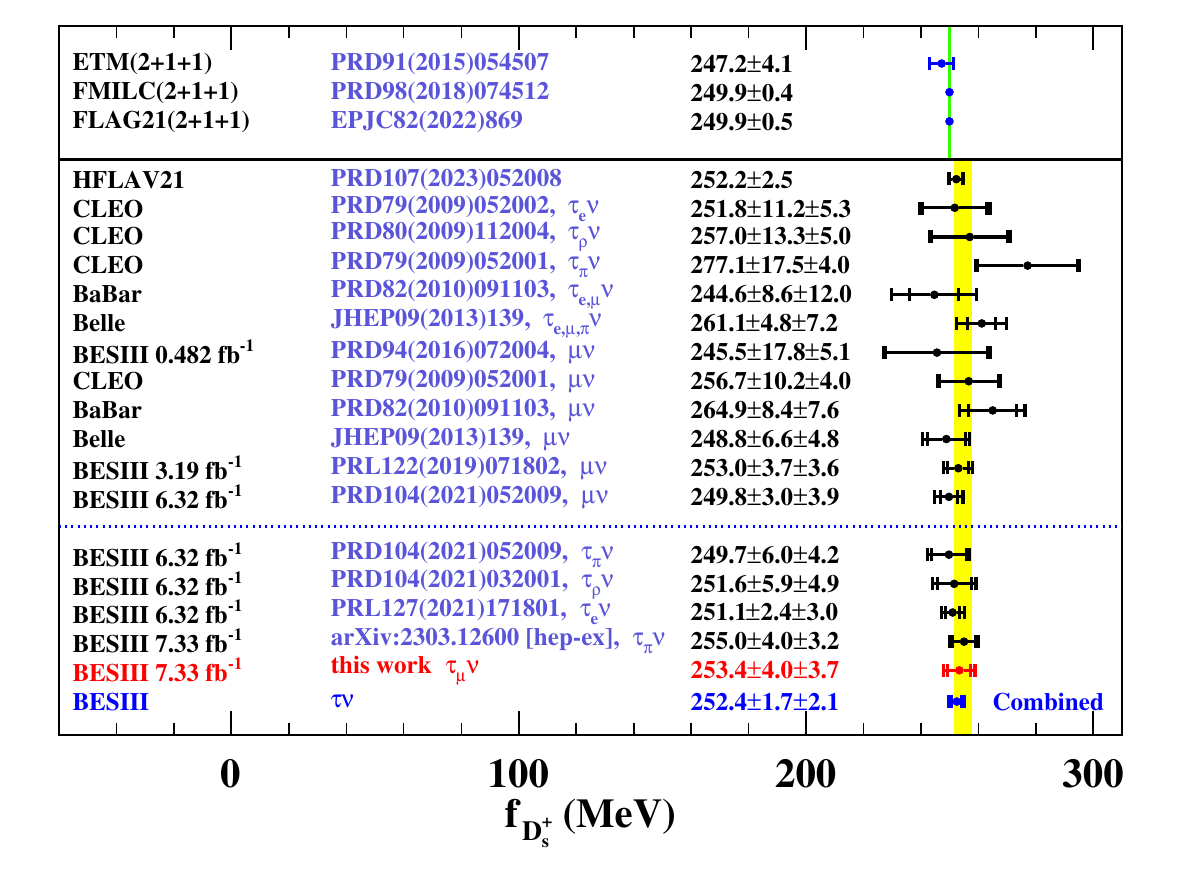}
\caption{Comparison of $f_{D^+_s}$ values in this with previous work and LQCD calculations. For experimental results, the inner error bar is the statistical uncertainty and the outer is the combined statistical and systematic uncertainty. The green band denotes the FLAG average and the yellow one denotes the experimental average. The last line is the BESIII combined result which does not include the BESIII result in Ref.~\cite{hajime2021}.}
\label{fig:comfds}
\end{figure}

\begin{figure}[htpb]\centering
\includegraphics[width=0.7\textwidth]{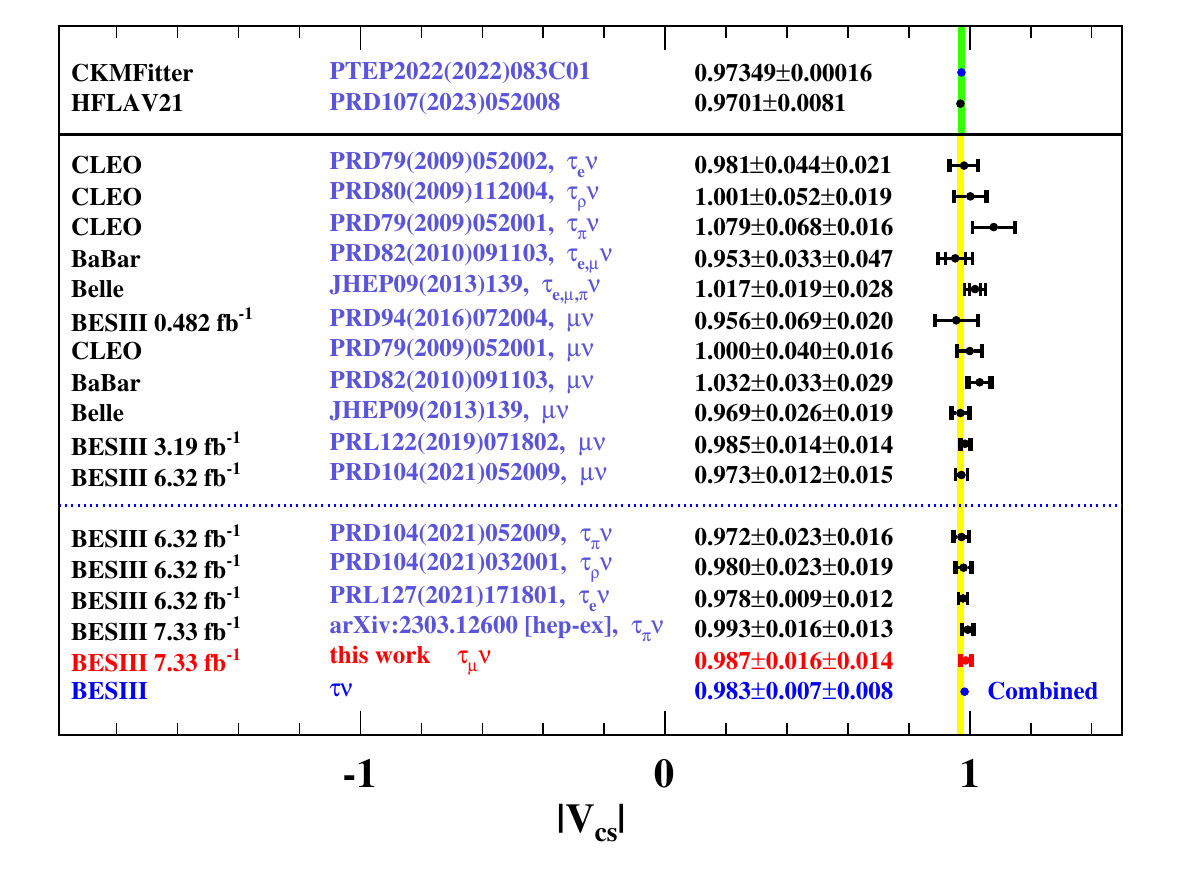}
\caption{Comparison of $|V_{\rm cs}|$ values in this with previous work. For experimental results, the inner error bar is the statistical uncertainty  and the outer is the combined statistical and systematic uncertainty. The green band denotes the CKM Fitter average and the yellow one denotes the experimental average. The last line is the  BESIII combined result which does not include the BESIII result in Ref.~\cite{hajime2021}.
}
\label{fig:comvcs}
\end{figure}

We determine an average~\cite{Schmelling:1994pz} BF for $D_s^+\to\tau^+\nu_\tau$ and the derived quantities that follow from this result,
taking as input the  BF measurement  from the current study, and those BF measurements using the decays  
$\tau^+\to\pi^+\pi^0\bar\nu_\tau$~\cite{tauvyue},
$\tau^+\to e^+\nu_e\bar\nu_\tau$~\cite{huijingevv} and
$\tau^+\to\pi^+\bar\nu_\tau$~\cite{xiechen}.
 The uncertainties from the ST yield, the $\pi^+$ tracking efficiency, the soft $\gamma(\pi^0)$ reconstruction, the best transition $\gamma(\pi^0)$ selection, the tag bias, $\tau_{D_s^+}$, $m_{D^+_s}$, $m_\tau$  and $|V_{cs}|$  are taken to be correlated between the measurements.
We determine  the average BF to be 
$\mathcal B(D_s^+\to\tau^+\nu_\tau)=(5.33\pm0.07_{\rm stat}\pm0.08_{\rm syst})\%$.
From this result it follows
$f_{D^+_s}=(252.4\pm1.7_{\rm stat}\pm2.1_{\rm syst})$~MeV,
$|V_{cs}|=0.983\pm0.007_{\rm stat}\pm0.008_{\rm syst}$, and
$\mathcal R_{\tau/\mu}=9.82\pm0.33$, again consistent with the expectation based on the assumption lepton flavor universality.
Figures~\ref{fig:combf}, \ref{fig:comfds}, and \ref{fig:comvcs} show comparisons of our results for ${\mathcal B}(D_s^+ \to \tau^+ \nu_\tau)$, $f_{D^+_s}$, and $|V_{cs}|$ with those of previous results.

Improved measurements of $\mathcal B(D_s^+\to\tau^+\nu_\tau)$ are foreseen with the larger data sets that BESIII is expected to accumulate in the coming years~\cite{white_paper}.

\section{ACKNOWLEDGEMENT}

The BESIII Collaboration thanks the staff of BEPCII and the IHEP computing center for their strong support. This work is supported in part by National Key R\&D Program of China under Contracts Nos. 2020YFA0406400, 2020YFA0406300; National Natural Science Foundation of China (NSFC) under Contracts Nos. 11875170, 12105076, 11635010, 11735014, 11835012, 11935015, 11935016, 11935018, 11961141012, 12022510, 12025502, 12035009, 12035013, 12061131003, 12192260, 12192261, 12192262, 12192263, 12192264, 12192265; the Chinese Academy of Sciences (CAS) Large-Scale Scientific Facility Program; the CAS Center for Excellence in Particle Physics (CCEPP); Joint Large-Scale Scientific Facility Funds of the NSFC and CAS under Contract No. U1832207; CAS Key Research Program of Frontier Sciences under Contracts Nos. QYZDJ-SSW-SLH003, QYZDJ-SSW-SLH040; 100 Talents Program of CAS; The Institute of Nuclear and Particle Physics (INPAC) and Shanghai Key Laboratory for Particle Physics and Cosmology; ERC under Contract No. 758462; European Union's Horizon 2020 research and innovation programme under Marie Sklodowska-Curie grant agreement under Contract No. 894790; German Research Foundation DFG under Contracts Nos. 443159800, 455635585, Collaborative Research Center CRC 1044, FOR5327, GRK 2149; Istituto Nazionale di Fisica Nucleare, Italy; Ministry of Development of Turkey under Contract No. DPT2006K-120470; National Research Foundation of Korea under Contract No. NRF-2022R1A2C1092335; National Science and Technology fund; National Science Research and Innovation Fund (NSRF) via the Program Management Unit for Human Resources \& Institutional Development, Research and Innovation under Contract No. B16F640076; Polish National Science Centre under Contract No. 2019/35/O/ST2/02907; Suranaree University of Technology (SUT), Thailand Science Research and Innovation (TSRI), and National Science Research and Innovation Fund (NSRF) under Contract No. 160355; The Royal Society, UK under Contract No. DH160214; The Swedish Research Council; U. S. Department of Energy under Contract No. DE-FG02-05ER41374.

\bibliographystyle{JHEP}
\bibliography{bibliography.bib}

\providecommand{\href}[2]{#2}\begingroup\raggedright\begin{thebibliography}{10}

\bibitem{decayrate}
D.~Silverman and H.~Yao, \emph{{Relativistic Treatment of Light Quarks in $D$
  and $B$ Mesons and $W$ Exchange Weak Decays}},
  \href{https://doi.org/10.1103/PhysRevD.38.214}{\emph{Phys. Rev. D} {\bfseries
  38} (1988) 214}.

\bibitem{FLab2018}
A.~Bazavov et~al., \emph{{$B$- and $D$-meson leptonic decay constants from
  four-flavor lattice QCD}},
  \href{https://doi.org/10.1103/PhysRevD.98.074512}{\emph{Phys. Rev. D}
  {\bfseries 98} (2018) 074512}
  [\href{https://arxiv.org/abs/1712.09262}{{\ttfamily 1712.09262}}].

\bibitem{LQCD}
{\scshape Fermilab Lattice, MILC} collaboration, \emph{{Charmed and Light
  Pseudoscalar Meson Decay Constants from Four-Flavor Lattice QCD with Physical
  Light Quarks}}, \href{https://doi.org/10.1103/PhysRevD.90.074509}{\emph{Phys.
  Rev. D} {\bfseries 90} (2014) 074509}
  [\href{https://arxiv.org/abs/1407.3772}{{\ttfamily 1407.3772}}].

\bibitem{etm2015}
N.~Carrasco et~al., \emph{{Leptonic decay constants $f_{K},f_{D},$ and
  $f_{{D}_{s}}$ with $N_{f} = 2+1+1$ twisted-mass lattice QCD}},
  \href{https://doi.org/10.1103/PhysRevD.91.054507}{\emph{Phys. Rev. D}
  {\bfseries 91} (2015) 054507}
  [\href{https://arxiv.org/abs/1411.7908}{{\ttfamily 1411.7908}}].

\bibitem{ukqcd2017}
P.A.~Boyle, L.~Del~Debbio, A.~J\"uttner, A.~Khamseh, F.~Sanfilippo and
  J.T.~Tsang, \emph{{The decay constants ${\mathbf{f_D}}$ and
  ${\mathbf{f_{D_{s}}}}$ in the continuum limit of ${\mathbf{N_f=2+1}}$ domain
  wall lattice QCD}},
  \href{https://doi.org/10.1007/JHEP12(2017)008}{\emph{JHEP} {\bfseries 12}
  (2017) 008} [\href{https://arxiv.org/abs/1701.02644}{{\ttfamily
  1701.02644}}].

\bibitem{ukqcd2015}
Y.-B.~Yang et~al., \emph{{Charm and strange quark masses and $f_{D_s}$ from
  overlap fermions}},
  \href{https://doi.org/10.1103/PhysRevD.92.034517}{\emph{Phys. Rev. D}
  {\bfseries 92} (2015) 034517}
  [\href{https://arxiv.org/abs/1410.3343}{{\ttfamily 1410.3343}}].

\bibitem{FLAG2021}
{\scshape Flavour Lattice Averaging Group (FLAG)} collaboration, \emph{{FLAG
  Review 2021}},
  \href{https://doi.org/10.1140/epjc/s10052-022-10536-1}{\emph{Eur. Phys. J. C}
  {\bfseries 82} (2022) 869}
  [\href{https://arxiv.org/abs/2111.09849}{{\ttfamily 2111.09849}}].

\bibitem{chen2014}
{\scshape TWQCD} collaboration, \emph{{Decay Constants of Pseudoscalar
  $D$-mesons in Lattice QCD with Domain-Wall Fermion}},
  \href{https://doi.org/10.1016/j.physletb.2014.07.025}{\emph{Phys. Lett. B}
  {\bfseries 736} (2014) 231}
  [\href{https://arxiv.org/abs/1404.3648}{{\ttfamily 1404.3648}}].

\bibitem{becirevic2013}
D.~Becirevic, B.~Blossier, A.~Gerardin, A.~Le~Yaouanc and F.~Sanfilippo,
  \emph{{On the significance of B-decays to radially excited D}},
  \href{https://doi.org/10.1016/j.nuclphysb.2013.04.008}{\emph{Nucl. Phys. B}
  {\bfseries 872} (2013) 313}
  [\href{https://arxiv.org/abs/1301.7336}{{\ttfamily 1301.7336}}].

\bibitem{wang2015}
Z.-G.~Wang, \emph{{Analysis of the masses and decay constants of the
  heavy-light mesons with QCD sum rules}},
  \href{https://doi.org/10.1140/epjc/s10052-015-3653-9}{\emph{Eur. Phys. J. C}
  {\bfseries 75} (2015) 427}
  [\href{https://arxiv.org/abs/1506.01993}{{\ttfamily 1506.01993}}].

\bibitem{cleo2009}
{\scshape CLEO} collaboration, \emph{{Measurement of $\mathcal{B}{D_s^+ \to
  \ell^+ \nu}$ and the Decay Constant $f_{D_s^+}$ From 600 $/pb^{-1}$ of
  $e^\pm$ Annihilation Data Near 4170 MeV}},
  \href{https://doi.org/10.1103/PhysRevD.79.052001}{\emph{Phys. Rev. D}
  {\bfseries 79} (2009) 052001}
  [\href{https://arxiv.org/abs/0901.1216}{{\ttfamily 0901.1216}}].

\bibitem{cleo2009a}
{\scshape CLEO} collaboration, \emph{{Measurement of the Pseudoscalar Decay
  Constant $f_{D_s^+}$ Using $D_s^+\to\tau^+\nu_\tau,
  \tau^+\to\rho^+\bar\nu_\tau$ Decays}},
  \href{https://doi.org/10.1103/PhysRevD.80.112004}{\emph{Phys. Rev. D}
  {\bfseries 80} (2009) 112004}
  [\href{https://arxiv.org/abs/0910.3602}{{\ttfamily 0910.3602}}].

\bibitem{cleo2009b}
{\scshape CLEO} collaboration, \emph{{Improved Measurement of Absolute
  Branching Fraction of $D_s^+\to\tau^+\nu_\tau$}},
  \href{https://doi.org/10.1103/PhysRevD.79.052002}{\emph{Phys. Rev. D}
  {\bfseries 79} (2009) 052002}
  [\href{https://arxiv.org/abs/0901.1147}{{\ttfamily 0901.1147}}].

\bibitem{babar2010}
{\scshape BaBar} collaboration, \emph{{Measurement of the Absolute Branching
  Fractions for $D^-_s\!\rightarrow\!\ell^-\bar{\nu}_{\ell}$ and Extraction of
  the Decay Constant $f_{D_s}$}},
  \href{https://doi.org/10.1103/PhysRevD.82.091103}{\emph{Phys. Rev. D}
  {\bfseries 82} (2010) 091103}
  [\href{https://arxiv.org/abs/1008.4080}{{\ttfamily 1008.4080}}].

\bibitem{belle2013}
{\scshape ATLAS} collaboration, \emph{{Searches for electroweak production of
  supersymmetric particles with compressed mass spectra in $\sqrt{s}=$ 13 TeV
  $pp$ collisions with the ATLAS detector}},
  \href{https://doi.org/10.1103/PhysRevD.101.052005}{\emph{Phys. Rev. D}
  {\bfseries 101} (2020) 052005}
  [\href{https://arxiv.org/abs/1911.12606}{{\ttfamily 1911.12606}}].

\bibitem{bes2016}
{\scshape BESIII} collaboration, \emph{{Measurement of the $D_s^+ \to
  \ell^+\nu_\ell$ branching fractions and the decay constant $f_{D_s^+}$}},
  \href{https://doi.org/10.1103/PhysRevD.94.072004}{\emph{Phys. Rev. D}
  {\bfseries 94} (2016) 072004}
  [\href{https://arxiv.org/abs/1608.06732}{{\ttfamily 1608.06732}}].

\bibitem{bes2019}
{\scshape BESIII} collaboration, \emph{{Determination of the pseudoscalar decay
  constant $f_{D_s^+}$ via $D_s^+\to\mu^+\nu_\mu$}},
  \href{https://doi.org/10.1103/PhysRevLett.122.071802}{\emph{Phys. Rev. Lett.}
  {\bfseries 122} (2019) 071802}
  [\href{https://arxiv.org/abs/1811.10890}{{\ttfamily 1811.10890}}].

\bibitem{hajime2021}
{\scshape BESIII} collaboration, \emph{{Measurement of the absolute branching
  fractions for purely leptonic $D_s^+$ decays}},
  \href{https://doi.org/10.1103/PhysRevD.104.052009}{\emph{Phys. Rev. D}
  {\bfseries 104} (2021) 052009}
  [\href{https://arxiv.org/abs/2102.11734}{{\ttfamily 2102.11734}}].

\bibitem{tauvyue}
{\scshape BESIII} collaboration, \emph{{Measurement of the branching fraction
  of leptonic decay $D_s^+\to\tau^+\nu_\tau$ via $\tau^+\to\pi^+\pi^0\bar
  \nu_\tau$}}, \href{https://doi.org/10.1103/PhysRevD.104.032001}{\emph{Phys.
  Rev. D} {\bfseries 104} (2021) 032001}
  [\href{https://arxiv.org/abs/2105.07178}{{\ttfamily 2105.07178}}].

\bibitem{huijingevv}
{\scshape BESIII} collaboration, \emph{{Measurement of the Absolute Branching
  Fraction of $D_s^+ \to \tau^+ \nu_{\tau}$ via $\tau^+ \to e^+ \nu_e
  \bar{\nu}_{\tau}$}},
  \href{https://doi.org/10.1103/PhysRevLett.127.171801}{\emph{Phys. Rev. Lett.}
  {\bfseries 127} (2021) 171801}
  [\href{https://arxiv.org/abs/2106.02218}{{\ttfamily 2106.02218}}].

\bibitem{xiechen}
{\scshape BESIII} collaboration, \emph{{Updated measurement of the branching
  fraction of $D_s^+\to \tau^+\nu_\tau$ via $\tau^+\to\pi^+\bar{\nu}_\tau$}},
  \href{https://arxiv.org/abs/2303.12600}{{\ttfamily 2303.12600}}.

\bibitem{haibo2021}
H.-B.~Li and X.-R.~Lyu, \emph{{Study of the standard model with weak decays of
  charmed hadrons at BESIII}},
  \href{https://doi.org/10.1093/nsr/nwab181}{\emph{Natl. Sci. Rev.} {\bfseries
  8} (2021) nwab181} [\href{https://arxiv.org/abs/2103.00908}{{\ttfamily
  2103.00908}}].

\bibitem{PDG2022}
{\scshape Particle Data Group} collaboration, \emph{{Review of Particle
  Physics}}, \href{https://doi.org/10.1093/ptep/ptaC097}{\emph{PTEP} {\bfseries
  2022} (2022) 083C01}.

\bibitem{BESIII:2015zbz}
{\scshape BESIII} collaboration, \emph{{Measurement of the center-of-mass
  energies at BESIII via the di-muon process}},
  \href{https://doi.org/10.1088/1674-1137/40/6/063001}{\emph{Chin. Phys. C}
  {\bfseries 40} (2016) 063001}
  [\href{https://arxiv.org/abs/1510.08654}{{\ttfamily 1510.08654}}].

\bibitem{ref_emc_energy}
{\scshape BESIII} collaboration, \emph{{Precision measurement of the integrated
  luminosity of the data taken by BESIII at center of mass energies between
  3.810 GeV and 4.600 GeV}},
  \href{https://doi.org/10.1088/1674-1137/39/9/093001}{\emph{Chin. Phys. C}
  {\bfseries 39} (2015) 093001}
  [\href{https://arxiv.org/abs/1503.03408}{{\ttfamily 1503.03408}}].

\bibitem{ref_emc_energy2}
{\scshape BESIII} collaboration, \emph{{Measurement of integrated luminosities
  at BESIII for data samples at center-of-mass energies between 4.0 and 4.6
  GeV}}, \href{https://doi.org/10.1088/1674-1137/ac80b4}{\emph{Chin. Phys. C}
  {\bfseries 46} (2022) 113002}
  [\href{https://arxiv.org/abs/2203.03133}{{\ttfamily 2203.03133}}].

\bibitem{Ablikim:2009aa}
{\scshape BESIII} collaboration, \emph{{Design and Construction of the BESIII
  Detector}}, \href{https://doi.org/10.1016/j.nima.2009.12.050}{\emph{Nucl.
  Instrum. Meth. A} {\bfseries 614} (2010) 345}
  [\href{https://arxiv.org/abs/0911.4960}{{\ttfamily 0911.4960}}].

\bibitem{Yu:IPAC2016-TUYA01}
C.~Yu et~al., \emph{{BEPCII Performance and Beam Dynamics Studies on
  Luminosity}},  in \emph{{7th International Particle Accelerator Conference}},
  p.~TUYA01, 2016, \href{https://doi.org/10.18429/JACoW-IPAC2016-TUYA01}{DOI}.

\bibitem{white_paper}
{\scshape BESIII} collaboration, \emph{{Future Physics Programme of BESIII}},
  \href{https://doi.org/10.1088/1674-1137/44/4/040001}{\emph{Chin. Phys. C}
  {\bfseries 44} (2020) 040001}
  [\href{https://arxiv.org/abs/1912.05983}{{\ttfamily 1912.05983}}].

\bibitem{youbes}
K.-X.~Huang, Z.-J.~Li, Z.~Qian, J.~Zhu, H.-Y.~Li, Y.-M.~Zhang et~al.,
  \emph{{Method for detector description transformation to Unity and
  application in BESIII}},
  \href{https://doi.org/10.1007/s41365-022-01133-8}{\emph{Nucl. Sci. Tech.}
  {\bfseries 33} (2022) 142}
  [\href{https://arxiv.org/abs/2206.10117}{{\ttfamily 2206.10117}}].

\bibitem{etofa}
X.~Li et~al., \emph{{Study of MRPC technology for BESIII endcap-TOF upgrade}},
  \href{https://doi.org/10.1007/s41605-017-0014-2}{\emph{Radiat Detect Technol
  Methods} {\bfseries 1} (2022) 12}.

\bibitem{etofb}
Y.~Guo et~al., \emph{{The study of time calibration for upgraded end cap TOF of
  BESIII}}, \href{https://doi.org/10.1007/s41605-017-0012-4}{\emph{Radiat
  Detect Technol Methods} {\bfseries 1} (2017) 14}.

\bibitem{etofc}
P.~Cao et~al., \emph{{Design and construction of the new BESIII endcap
  Time-of-Flight system with MRPC Technology}},
  \href{https://doi.org/10.1016/j.nima.2019.163053}{\emph{Nucl. Instrum. Meth.
  A} {\bfseries 953} (2020) 163053}.

\bibitem{geant4}
{\scshape GEANT4} collaboration, \emph{{GEANT4--a simulation toolkit}},
  \href{https://doi.org/10.1016/S0168-9002(03)01368-8}{\emph{Nucl. Instrum.
  Meth. A} {\bfseries 506} (2003) 250}.

\bibitem{ref:kkmc1}
S.~Jadach, B.F.L.~Ward and Z.~Was, \emph{{Coherent exclusive exponentiation for
  precision Monte Carlo calculations}},
  \href{https://doi.org/10.1103/PhysRevD.63.113009}{\emph{Phys. Rev. D}
  {\bfseries 63} (2001) 113009}
  [\href{https://arxiv.org/abs/hep-ph/0006359}{{\ttfamily hep-ph/0006359}}].

\bibitem{ref:kkmc2}
S.~Jadach, B.F.L.~Ward and Z.~Was, \emph{{The Precision Monte Carlo event
  generator K K for two fermion final states in e+ e- collisions}},
  \href{https://doi.org/10.1016/S0010-4655(00)00048-5}{\emph{Comput. Phys.
  Commun.} {\bfseries 130} (2000) 260}
  [\href{https://arxiv.org/abs/hep-ph/9912214}{{\ttfamily hep-ph/9912214}}].

\bibitem{ref:conexc}
R.-G.~Ping, \emph{{An exclusive event generator for $e^+ e^-$ scan
  experiments}},
  \href{https://doi.org/10.1088/1674-1137/38/8/083001}{\emph{Chin. Phys. C}
  {\bfseries 38} (2014) 083001}
  [\href{https://arxiv.org/abs/1309.3932}{{\ttfamily 1309.3932}}].

\bibitem{ref:evtgen1}
D.J.~Lange, \emph{{The EvtGen particle decay simulation package}},
  \href{https://doi.org/10.1016/S0168-9002(01)00089-4}{\emph{Nucl. Instrum.
  Meth. A} {\bfseries 462} (2001) 152}.

\bibitem{ref:evtgen2}
R.-G.~Ping, \emph{{Event generators at BESIII}},
  \href{https://doi.org/10.1088/1674-1137/32/8/001}{\emph{Chin. Phys. C}
  {\bfseries 32} (2008) 599}.

\bibitem{ref:lundcharm1}
J.C.~Chen, G.S.~Huang, X.R.~Qi, D.H.~Zhang and Y.S.~Zhu, \emph{{Event generator
  for J / psi and psi (2S) decay}},
  \href{https://doi.org/10.1103/PhysRevD.62.034003}{\emph{Phys. Rev. D}
  {\bfseries 62} (2000) 034003}.

\bibitem{ref:lundcharm2}
R.-L.~Yang, R.-G.~Ping and H.~Chen, \emph{{Tuning and Validation of the
  Lundcharm Model with $J/\psi$ Decays}},
  \href{https://doi.org/10.1088/0256-307X/31/6/061301}{\emph{Chin. Phys. Lett.}
  {\bfseries 31} (2014) 061301}.

\bibitem{photos}
E.~Richter-Was, \emph{{QED bremsstrahlung in semileptonic B and leptonic tau
  decays}}, \href{https://doi.org/10.1016/0370-2693(93)90062-M}{\emph{Phys.
  Lett. B} {\bfseries 303} (1993) 163}.

\bibitem{DTmethod1}
{\scshape MARK-III} collaboration, \emph{{Direct Measurements of Charmed d
  Meson Hadronic Branching Fractions}},
  \href{https://doi.org/10.1103/PhysRevLett.56.2140}{\emph{Phys. Rev. Lett.}
  {\bfseries 56} (1986) 2140}.

\bibitem{Schmelling:1994pz}
M.~Schmelling, \emph{{Averaging correlated data}},
  \href{https://doi.org/10.1088/0031-8949/51/6/002}{\emph{Phys. Scripta}
  {\bfseries 51} (1995) 676}.

\bibitem{bes3_etaev}
{\scshape BESIII} collaboration, \emph{{Measurement of the Dynamics of the
  Decays $D_s^+ \rightarrow \eta^{(\prime)} e^+ \nu_e$}},
  \href{https://doi.org/10.1103/PhysRevLett.122.121801}{\emph{Phys. Rev. Lett.}
  {\bfseries 122} (2019) 121801}
  [\href{https://arxiv.org/abs/1901.02133}{{\ttfamily 1901.02133}}].

\bibitem{bes3_gev}
{\scshape BESIII} collaboration, \emph{{Search for the decay $D_s^+\rightarrow
  \gamma e^+\nu_e$}},
  \href{https://doi.org/10.1103/PhysRevD.99.072002}{\emph{Phys. Rev. D}
  {\bfseries 99} (2019) 072002}
  [\href{https://arxiv.org/abs/1902.03351}{{\ttfamily 1902.03351}}].

\bibitem{cleo_dstotaunu_enunu1}
{\scshape CLEO} collaboration, \emph{{Measurement of the absolute branching
  fraction of $D_s^+\to\tau^+\nu_\tau$ decay}},
  \href{https://doi.org/10.1103/PhysRevLett.100.161801}{\emph{Phys. Rev. Lett.}
  {\bfseries 100} (2008) 161801}
  [\href{https://arxiv.org/abs/0712.1175}{{\ttfamily 0712.1175}}].

\bibitem{cleo_dstotaunu_enunu2}
{\scshape CLEO} collaboration, \emph{{Improved Measurement of Absolute
  Branching Fraction of $D_s^+\to\tau^+\nu_\tau$}},
  \href{https://doi.org/10.1103/PhysRevD.79.052002}{\emph{Phys. Rev. D}
  {\bfseries 79} (2009) 052002}
  [\href{https://arxiv.org/abs/0901.1147}{{\ttfamily 0901.1147}}].

\bibitem{luminosity_3773}
{\scshape BESIII} collaboration, \emph{{Measurement of the integrated
  luminosities of the data taken by BESIII at $\sqrt{s}=$3.650 and 3.773 GeV}},
  \href{https://doi.org/10.1088/1674-1137/37/12/123001}{\emph{Chin. Phys. C}
  {\bfseries 37} (2013) 123001}
  [\href{https://arxiv.org/abs/1307.2022}{{\ttfamily 1307.2022}}].

\bibitem{luminosity_3773a}
{\scshape BESIII} collaboration, \emph{{Measurement of the $e^+ e^- \to \pi^+
  \pi^-$ cross section between 600 and 900 MeV using initial state radiation}},
  \href{https://doi.org/10.1016/j.physletb.2015.11.043}{\emph{Phys. Lett. B}
  {\bfseries 753} (2016) 629}
  [\href{https://arxiv.org/abs/1507.08188}{{\ttfamily 1507.08188}}].

\bibitem{geff}
{\scshape BESIII} collaboration, \emph{{Study of $\chi_{cJ}$ radiative decays
  into a vector meson}},
  \href{https://doi.org/10.1103/PhysRevD.83.112005}{\emph{Phys. Rev. D}
  {\bfseries 83} (2011) 112005}
  [\href{https://arxiv.org/abs/1103.5564}{{\ttfamily 1103.5564}}].

\bibitem{pi0eff}
{\scshape BESIII} collaboration, \emph{{Observation of the $W$-Annihilation
  Decay $D^{+}_{s} \rightarrow \omega \pi^{+}$ and Evidence for $D^{+}_{s}
  \rightarrow \omega K^{+}$}},
  \href{https://doi.org/10.1103/PhysRevD.99.091101}{\emph{Phys. Rev. D}
  {\bfseries 99} (2019) 091101}
  [\href{https://arxiv.org/abs/1811.00392}{{\ttfamily 1811.00392}}].

\bibitem{ckmfitter}
{\scshape CKMfitter Group} collaboration, \emph{{CP violation and the CKM
  matrix: Assessing the impact of the asymmetric $B$ factories}},
  \href{https://doi.org/10.1140/epjc/s2005-02169-1}{\emph{Eur. Phys. J. C}
  {\bfseries 41} (2005) 1}
  [\href{https://arxiv.org/abs/hep-ph/0406184}{{\ttfamily hep-ph/0406184}}].

\bibitem{bes3_kev}
{\scshape BESIII} collaboration, \emph{{Study of Dynamics of $D^0 \to K^- e^+
  \nu_{e}$ and $D^0\to\pi^- e^+ \nu_{e}$ Decays}},
  \href{https://doi.org/10.1103/PhysRevD.92.072012}{\emph{Phys. Rev. D}
  {\bfseries 92} (2015) 072012}
  [\href{https://arxiv.org/abs/1508.07560}{{\ttfamily 1508.07560}}].

\bibitem{bes3_ksev}
{\scshape BESIII} collaboration, \emph{{Analysis of $D^+\to\bar K^0e^+\nu_e$
  and $D^+\to\pi^0e^+\nu_e$ semileptonic decays}},
  \href{https://doi.org/10.1103/PhysRevD.96.012002}{\emph{Phys. Rev. D}
  {\bfseries 96} (2017) 012002}
  [\href{https://arxiv.org/abs/1703.09084}{{\ttfamily 1703.09084}}].

\bibitem{bes3_klev}
{\scshape BESIII} collaboration, \emph{{Study of decay dynamics and $CP$
  asymmetry in $D^+ \to K^0_L e^+ \nu_e$ decay}},
  \href{https://doi.org/10.1103/PhysRevD.92.112008}{\emph{Phys. Rev. D}
  {\bfseries 92} (2015) 112008}
  [\href{https://arxiv.org/abs/1510.00308}{{\ttfamily 1510.00308}}].

\bibitem{bes3_kmuv}
{\scshape BESIII} collaboration, \emph{{Study of the $D^0 \rightarrow K^- \mu^+
  \nu_\mu$ Dynamics and Test of Lepton Flavor Universality with $D^0
  \rightarrow K^- \ell^+ \nu_\ell$ Decays}},
  \href{https://doi.org/10.1142/9789811217739_0041}{\emph{Phys. Rev. Lett}
  {\bfseries 122} (2019) 011804}.

\end{thebibliography}\endgroup

\end{document}